\documentclass{llncs}
\pdfoutput=1
\usepackage{fullpage,amsmath, amssymb, subfigure, enumerate, graphicx}

\newtheorem{obs}{Observation}
\newcommand{\wrb}{\omega} 
\newcommand{\leftn}{left}
\newcommand{\rightn}{right}
\newcommand{\HH}{{\mathcal H}} 
\newcommand{\C}{{\mathcal C}}

\title{Highway Hull Revisited}

\author{Greg~Aloupis\inst{1} 
\and Jean~Cardinal\inst{1} \and
S\'ebastien~Collette\thanks{Charg\'e de Recherches du F.R.S.-FNRS.}\inst{1} 
\and Ferran~Hurtado\inst{2}
\and Stefan~Langerman\thanks{Ma\^itre de Recherches du F.R.S.-FNRS.}\inst{1} \and
Joseph~O'Rourke\inst{3}
\and Bel\'en~Palop\inst{4}}

\institute{
Universit\'e Libre de Bruxelles (ULB), CP212,
Bld. du Triomphe, 1050 Brussels, Belgium.\\
E-mail: \email{\{greg.aloupis,jcardin,secollet,slanger\}@ulb.ac.be}
\and
Universitat Polit\`ecnica de Catalunya, Jordi Girona 1--3, E-08034 Barcelona, Spain.\\
E-mail: \email{Ferran.Hurtado@upc.edu}
\and
Smith College, Northampton, MA 01063, USA.\\
E-mail: \email{orourke@cs.smith.edu}
\and
Universidad de Valladolid, Spain.\\
E-mail: \email{b.palop@infor.uva.es}
}

\begin{document}
\maketitle
\sloppy

\begin{abstract}
A highway $H$ is a line in the plane on which one can travel at a greater speed than in the remaining plane.
One can choose to enter and exit $H$ at any point.
 The \emph{highway time distance} between a pair of points is the minimum time required to move from one point to the other, with optional use of $H$. 

\hspace{5mm}
The \emph{highway hull} $\HH(S,H)$ of a point set $S$ 
is the minimal set containing $S$ as well as the shortest paths between all pairs of points in $\HH(S,H)$, using the highway time distance.

\hspace{5mm}
We provide a $\Theta(n \log n)$ worst-case time algorithm to find the highway hull 
under the $L_{1}$ metric, as well as an $O(n \log^2 n)$ time algorithm for the $L_2$ metric which
improves the best known result of $O(n^2)$~\cite{HPS99,P03}. 
We also define and construct the \emph{useful region} of the plane:
the region that a highway must intersect in order that the shortest path between at least one pair of points uses the highway.\end{abstract}
\section{Introduction}

In recent years, much work has been done on geometric problems that stem from
Geographic Information Systems or involve transportation networks. Various geometric models 
of transportation have been proposed, and fundamental problems incorporating
these models have been analyzed, such as Voronoi diagrams and facility location. 

A simple model of transportation in the plane is that of the {\em highway}, defined as a line on which 
one can move at some speed $v>1$, while the speed off the highway is $1$. One can enter and exit the highway
at any point, and the distance between two points is defined as the minimum time to get from one to the other,
using the highway or not. Thus the shortest path between two points is either the line segment between
them, or a three-part piecewise linear path, the middle segment of which lies on the highway.

A natural notion to explore, first defined by Hurtado, Palop, and Sacrist\'an~\cite{HPS99}, is that of the convex hull 
in the presence of a highway. They define convexity with respect to the above definition of a shortest path. 

Hence a set $S$ will be convex in that sense if the shortest path between two points of $S$ is contained in the set.
We define the {\em highway hull} $\HH(S,H)$ of a set $S$ and highway $H$ as the closure of $S$ with respect to the inclusion of shortest paths. In other words, $\HH(S,H)$ is defined recursively. When a new point $x$ belonging to a shortest path is added to $\HH(S,H)$, we must also consider all shortest paths from $x$ to other points in $\HH(S,H)$.
This yields a minimal region containing the points and all shortest paths between any two points of 
the region. $\HH(S,H)$ is known~\cite{P03} to be decomposable into convex pieces that partition the point set into 
``clusters'' along the highway. It is therefore a simple tool for exploring the formation of communities along a straight path.

We present algorithms for computing the highway hull of $n$ points in subquadratic time
in two different distance metrics.
We first 
define the {\em orthogonal highway hull} using $L_1$ geometry, and provide a simple incremental 
 $O(n\log n)$ time algorithm. 
In the Euclidean case, we show that crucial properties of the orthogonal highway hull fail to hold. We provide an algorithm to compute the Euclidean highway hull in $O(n\log^2 n)$ time. 
Finally we propose an $O(n\log n)$ time algorithm for deciding whether a highway is useful for some 
pair of points in a given set. This involves computing the ``useful region'' of the point set, such that a 
highway is used if and only if it intersects this region.

\subsection{Related Work}
The notion of convex hull in the presence of a highway was recently studied by Yu and Lee~\cite{YL07},
who provided an incremental algorithm similar to ours. Unfortunately it does not yield a correct answer in all circumstances, because a number of critical cases that make the problem more difficult were overlooked. We give a precise
description of those cases in Section~\ref{sec:euclidean}, and explain how we can take them into account in the 
incremental algorithm. 
The only previous correct algorithm we are aware of was proposed by Palop~\cite{P03}, and constructs the Euclidean highway hull in $O(n^2)$ time.
\newline

Properties of several fundamental computational geometry structures within the presence of a transportation network have been analyzed (e.g.,~Voronoi diagrams \cite{AHIK03,AHP04}, skeletons~\cite{AAP04}). A systematic study on this topic can be found in~\cite{P03}, which contains an analysis of convex hulls, disks and wedges, as well as the aforementioned results.
Other related works include~\cite{LLW04,BC04,GSW07}. 

A common issue in the contributions mentioned above is the use of time metrics, where the distance between two points is a function of the time required to go from one point to the other. The model of highways defined above is one example, and variants can be found in the literature (e.g.,~\cite{AHP05,CL06,CCHLP07c}).

Recently, efforts have been made~\cite{CL06,CCHLP07c,AAA07} to solve the problem of optimally positioning transportation devices (highways, walkways, elevators, \emph{etc}.) in order to minimize the maximum travel time among a set of points. 
For instance in~\cite{AAA07}, algorithms are given to optimally place a highway, 
when the travel speed on the highway is infinite, or when the highway is restricted to be vertical. Moreover, an $O(n^2\log n)$ algorithm for the general problem is given. 
Various other highway models are studied in
\cite{KT07,KT08}.

Finally, other questions related to highways are studied in~\cite{CLLP05}. For example, in \emph{highway pricing games}, the problem is to define a price that a customer has to pay to gain access to some part of a transportation network. If the price on some path is too high, the customer might choose another route; the goal is to maximize the total price paid by customers.

\section{The Orthogonal Highway Hull}
\label{sec:ortho}

\subsection{Model}
Let the highway $H$ be the $x$-axis. Thus we can abbreviate our notation for the highway hull to $\HH(S)$. Even without a highway, the shortest path in $L_{1}$ is not necessarily unique. Given two points $p$ and $q$ with different $x$ and $y$-coordinates, an infinite number of shortest paths exist, all contained in the bounding box of $p$ and $q$. We will always choose the L-shaped shortest path whose corner is closest to $H$. Of course, this is just a convention, but uniqueness is useful in solving this problem.

The input is a set $S$ of $n$ points and a real number $v>1$.
For any point $p \in S$, let $p'$ be its orthogonal projection onto $H$.

\noindent The shortest path between two points $a$ and $b$ consists of either:
\begin{itemize}
\item the $L_1$ segment between $a$ and $b$ oriented toward $H$, or
\item the horizontal segment $a'b'$ and the vertical segments $aa'$ and $bb'$.
\end{itemize}

\begin{figure}
\begin{center}
\subfigure[Shortest paths.]{\includegraphics[angle=-90, scale=.5]{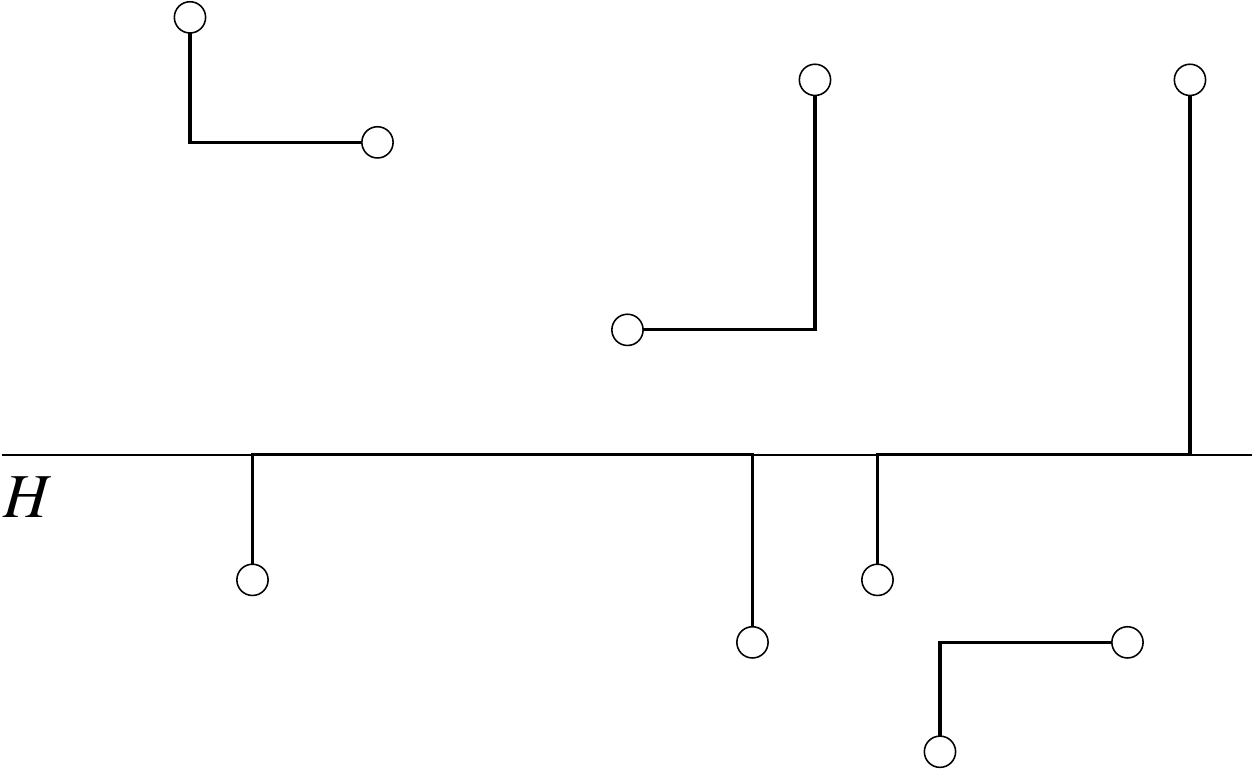}}
\hspace{1cm}
\subfigure[The orthogonal highway hull.]{\includegraphics[angle=-90, scale=.5]{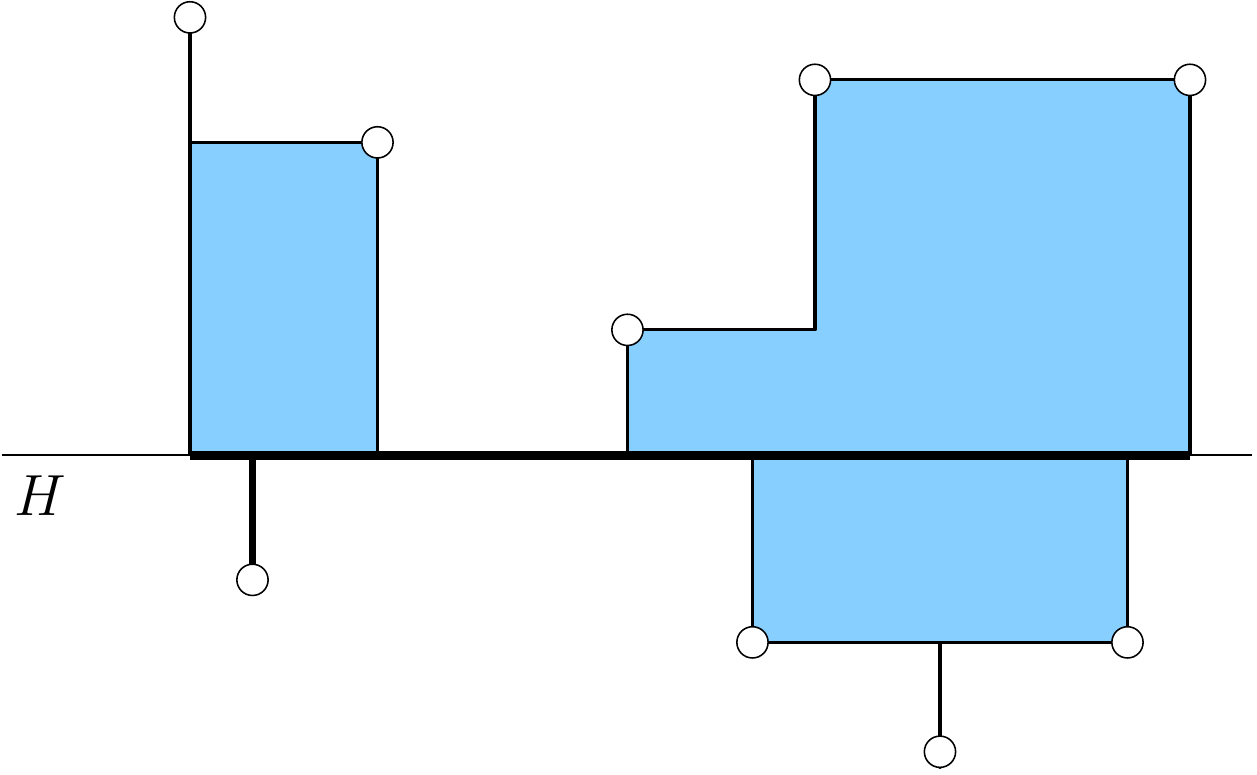}}
\end{center}
\caption{\label{fig:sp}Illustration of the definitions of shortest paths and orthogonal highway hull for $v=+\infty$.}
\end{figure}

In the first case the distance is the sum of the lengths of the two segments
(i.e.,~the $L_1$ distance between $a$ and $b$). In the latter case it is the sum $|aa'| + |a'b'|/v + |bb'|$, where $|xy|$ denotes the length of the segment $xy$. 
If the two distances are equal, we assume that $H$ is
not used. These conventions render the shortest path between two points unique. Examples of shortest paths are given in Figure~\ref{fig:sp}.

The orthogonal highway hull $\HH_1(S)$ of a point set $S$ and a horizontal line $H$ is the closure of $S$ under the operation of
including the orthogonal shortest path between each pair of points. It is therefore a minimal region containing the shortest paths
between any pair of points in the region.

\subsection{Incremental Algorithm: Preliminary Observations and Lemmas}

We start with a number of simple observations, inspired by similar observations for the $L_{2}$ case from~\cite{P03,YL07}.
First, note that $\HH_1(S)$ has no holes. The proof is by contradiction: without loss of generality, suppose there is a hole $h$ in $\HH_1(S)$ above $H$. Then, one can pick two points on the boundary of $h$ with same $x$-coordinate. The shortest path between these two points is then the vertical segment between them, which is contained inside $h$, a contradiction.

If at least one pair of points uses the highway, $\HH_1(S)$ can be decomposed into two parts: one segment along the highway, and a set of $L_{1}$-convex connected components, which we call
{\em clusters}. Since clusters are disjoint, a path between two clusters will always use $H$. 

If no pair of points use $H$, then the upper orthogonal hull is similar to the classical orthogonal 
convex hull (see for instance~\cite{OSW84}), while the lower orthogonal hull is the lower part of the bounding box of the the point set. 
We will rule out this case for now, and assume that at least one pair of points uses $H$. At the end of this section, we will show that our algorithm can be extended to also test if at least one pair uses the highway. If this is not the case, we simply compute the upper orthogonal hull in $\Theta(n \log n)$ time.

Another observation is that if the shortest path between two points $a$ and $b$ in $S$
does not use $H$, then all points that have an orthogonal projection between $a'$ and $b'$ 
belong to the same cluster.
	Indeed, the shortest path from every point $x$ on the segment $ab$ to $x'$ does not use $H$. Thus every point below $ab$ and above $H$ must belong to one cluster. Let $y$ be a point above $ab$, and $a' \prec_{x} y'  \prec_{x} b'$, where $a' \prec_{x} y'$ denotes that the $x$-coordinate of $a'$ is less than that of $y'$. The shortest path from $y$ to $y'$ intersects $ab$, which implies that $y$ belongs to the same cluster.

If at least one pair of points uses $H$, then the highway hull contains the projections of all $n$ points on $H$. This happens because at least one point $x$ of the hull is on $H$, and the shortest path from $x$ to any point $y\in S$ goes through $y'$.

Without loss of generality, and in order to simplify the exposition, we will suppose that all points are above $H$ and that at least one pair uses it.
We give an incremental algorithm for constructing the orthogonal highway hull, similar to the well-known Graham scan~\cite{Graham}. In a preprocessing
step, all points are sorted along the $x$ axis. Then each point is considered successively in sorted order.
Our algorithm is similar to the that of Yu and Lee~\cite{YL07} for the Euclidean case.

\begin{figure}
\begin{center}
\includegraphics[angle=-90, scale=.5]{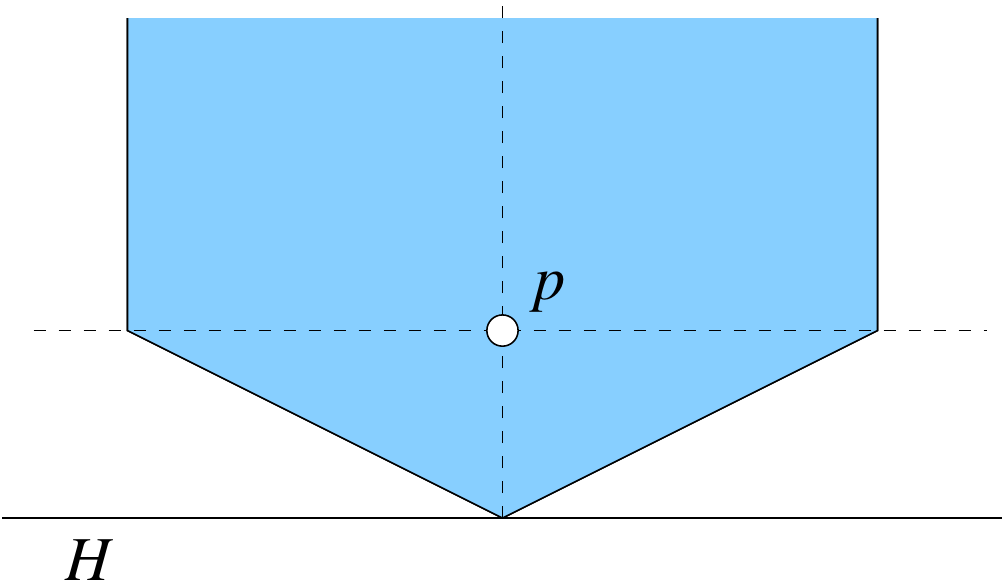}
\end{center}
\caption{\label{fig:wr}Walking region of a point $p$.}
\end{figure}

The {\em walking region} of a point $p$ is the set of points $q$ such that the shortest path between $p$ and $q$ does not use the 
highway (see Figure~\ref{fig:wr}). 
\noindent Hence it is the set $\{q \in\mathbb{R}^2 : |pq| \leq |pp'|+|p'q'|/v+|qq'|\}$.
In the orthogonal setting, the boundary of a walking region is composed of two line segments and two halflines. The two segments meet at $p'$ and join the halflines at the horizontal line through $p$. The segments have slopes $(1-1/v)/2$ and $-(1-1/v)/2$ respectively, independent of the position of $p$. We denote by $\wrb (p)$ the right boundary of the walking region of point $p$.

Similarly, we define the walking region of a set of points as the union of the walking regions of the points. 
We denote by $\wrb (\C)$ 
the right boundary of the walking region of a cluster $\mathcal C$. In general, the union of the walking regions of a set of points might contain holes. We only take into account the rightmost component of the boundary of the union of the walking regions. 

\begin{figure}
\begin{center}
\includegraphics[angle=-90,scale=.5]{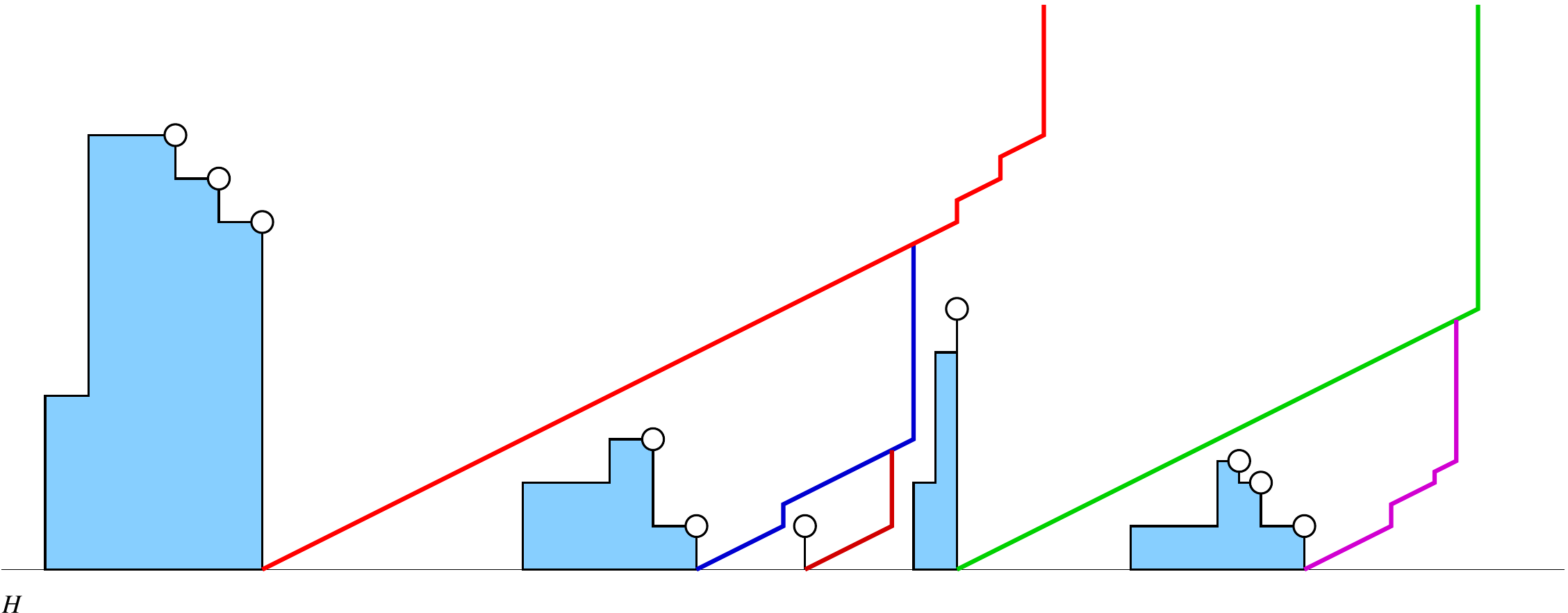}
\end{center}
\caption{\label{fig:wrcluster}Right boundaries of the walking regions of a set of clusters.}
\end{figure}

\noindent A simple property of these walking regions allows us to simplify the problem:
\begin{obs}
\label{lem:vtx}
If $a$ is above and to the right of $b$, then $\wrb (\{ a,b\}) = \wrb (a)$.
\end{obs}
As a consequence, the walking region of a cluster is the union of the walking regions of all convex vertices on the right side of the cluster.
This is illustrated in Figure~\ref{fig:wrcluster}.

Consider an orthogonal highway hull and the corresponding partition of the points into clusters. The clusters $\C_i$ are indexed
from left to right along $H$.
\begin{lemma}
\label{lem:intersection}
The right boundaries of the walking regions of two clusters $\C_i$ and $\C_j$, with $i<j$, can only intersect once.
Furthermore, the intersection point lies on the vertical halfline of $\wrb (\C_j)$.
\end{lemma}
\begin{proof}
Since the segments of the walking regions are parallel to each other, the intersection between two walking regions always takes place between a segment and a halfline. If there is an intersection between $\wrb (\C_i)$ and $\wrb (\C_j)$, then $\C_j$ must be to the left of the rightmost vertical halfline of $\wrb (\C_i)$. Since they are different clusters, $\C_j$ is entirely below $\wrb (\C_i)$. 
Thus the walking region of each point in $\C_j$ intersects $\wrb (\C_i)$ with its vertical half-line, and so does $\wrb (\C_j)$.
\qed\end{proof}

\begin{lemma}
\label{lem:order}
Consider a vertical ray $r$ emanating upward from a point $p$. The ray crosses the boundaries $\wrb (\C_i)$ in 
the right-to-left order of the clusters.
\end{lemma}
\begin{proof}
We can ignore the vertical halflines of the walking region boundaries since they are parallel to $r$. We prove that $r$ crosses sloped segments corresponding to the clusters in right-to-left order. 

Notice that all sloped segments are parallel (their slope only depends on the speed $v$, which is fixed). Each segment is defined by one point $x$ of the set, and the line though it crosses $H$ at $x'$. As the lines are parallel, they have the same sorted order as their corresponding points, i.e.,~the leftmost point corresponds to the topmost line. 

The ray crosses all the lines defined above in right-to-left order of the corresponding points. This implies that $r$ touches a subset of the segments, in the same order.

Finally, note that each cluster corresponds to an interval in the $x$-sorted order of the points, meaning that the order of the points corresponds to the order the the clusters containing them. 
\qed\end{proof}

\subsection{Incremental Algorithm: Description}

We focus on the task of partitioning the points into clusters $\{ \C_i\}$.
We label the points $p_1,p_2,\ldots ,p_n$ in increasing order of their $x$-coordinate. At the $i$th step, the algorithm
considers $p_i$. We decide whether $p_i$ will be included in an existing cluster. If not, we 
create a new cluster consisting of a vertical segment between $p_i$ and $p'_i$.

 $\wrb (\C_i)$ can be represented by a linked list $L_i$ of segments 
with slope $(1-1/v)/2$ (we do not need to store the vertical segments). We maintain a set $\mathcal L$ of non-empty lists, each associated with a cluster, representing boundaries that have not yet been entirely scanned. Suppose that the rightmost cluster is $\C_j$. The algorithm starts by scanning the 
list $L_j$ from left to right. At each step, we consider the relative position of $p_i$ to the current line 
segment $s$ in $L_j$. Let $s'$ be the projection of $s$ onto $H$.\bigskip

\noindent Four cases can occur:
\begin{enumerate}
\item If $p'_i$ is to the right of $s'$, we cannot decide yet whether $p_i$ is above or below $\wrb (\C_j)$. Thus we advance to the next segment in $L_j$.
The segment $s$ will never be used again and can be deleted from $L_j$, since all the other points are to the right of $p_i$. 

\item \label{case:stop} If $p_i$ is below $s$, 
we know from Lemma~\ref{lem:order} that $p_i$ is also outside the boundary of the walking region of any cluster to the left of $\C_j$.
Thus
$p_i$ forms its own cluster. In this case, an entire prefix of $L_j$ will be deleted.

\item If no segments remain in $L_j$, it still may be the case that $p_i$ belongs to another cluster to the left of $\C_j$. Therefore we start over with $L_{j-1}$ and remove $L_j$ from $\mathcal L$.

\item \label{case:iter} 
If $p_i$ lies above $s$, it must be merged with $\C_j$. However, $p_i$ might also need to merge with a 
cluster that is further to the left of $\C_j$. We must identify which is the {\em leftmost} cluster with which $p_i$ will merge. Similar to the previous case, we start over with $L_{j-1}$.
\end{enumerate}

Once we have deleted the prefixes of the lists corresponding to all the clusters that we have to merge, it remains to update the list structure. 
We create a list for the new cluster, that is composed of the points of $k$ previous clusters. Denote by $\mathcal C$ the leftmost cluster with which $p_i$ is merged. 
\begin{lemma}\label{lem:inbetween}
Let $x$ be any convex vertex of $\mathcal C$ whose
walking region contains $p_i$. Then all the clusters between $\mathcal C$ and $p_i$ are below the shortest path from $p_i$ to $x$.
\end{lemma}
\begin{proof}
Let $t$ be the shortest path between $p_i$ and $x$, and suppose that $t$ intersects two distinct clusters, at two points $a$ and $b$. By definition, $t$ does not use $H$. A subpath of a shortest path is also a shortest path, which means that the shortest path
between $a$ and $b$ does not use $H$ either. Hence $a$ and $b$ should belong to the same cluster, a contradiction.\qed
\end{proof}

\begin{figure}
\subfigure[\label{fig:before}Before]{\includegraphics[angle=-90,scale=.35]{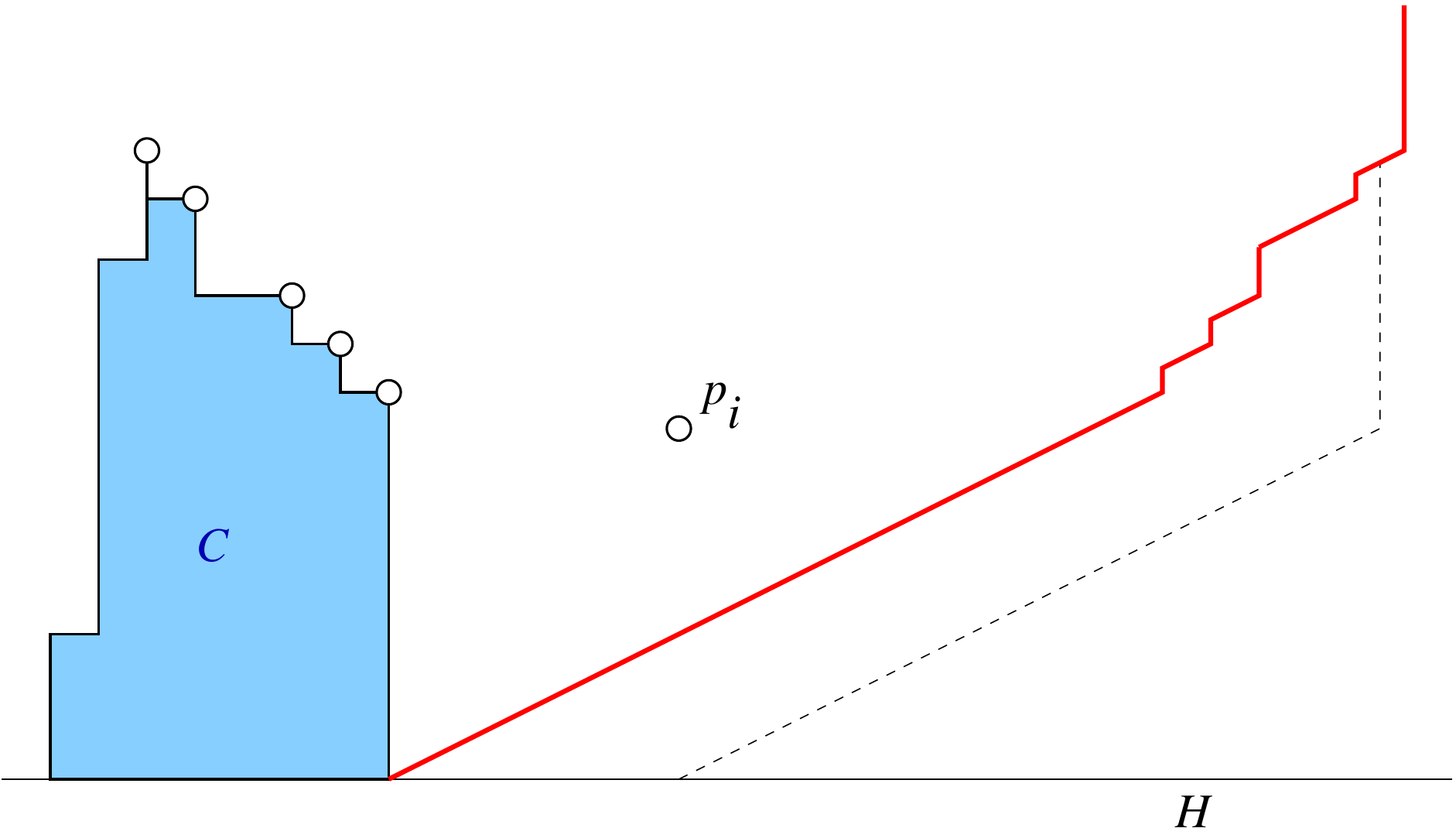}}
\hspace{1cm}
\subfigure[\label{fig:after}After]{\includegraphics[angle=-90,scale=.35]{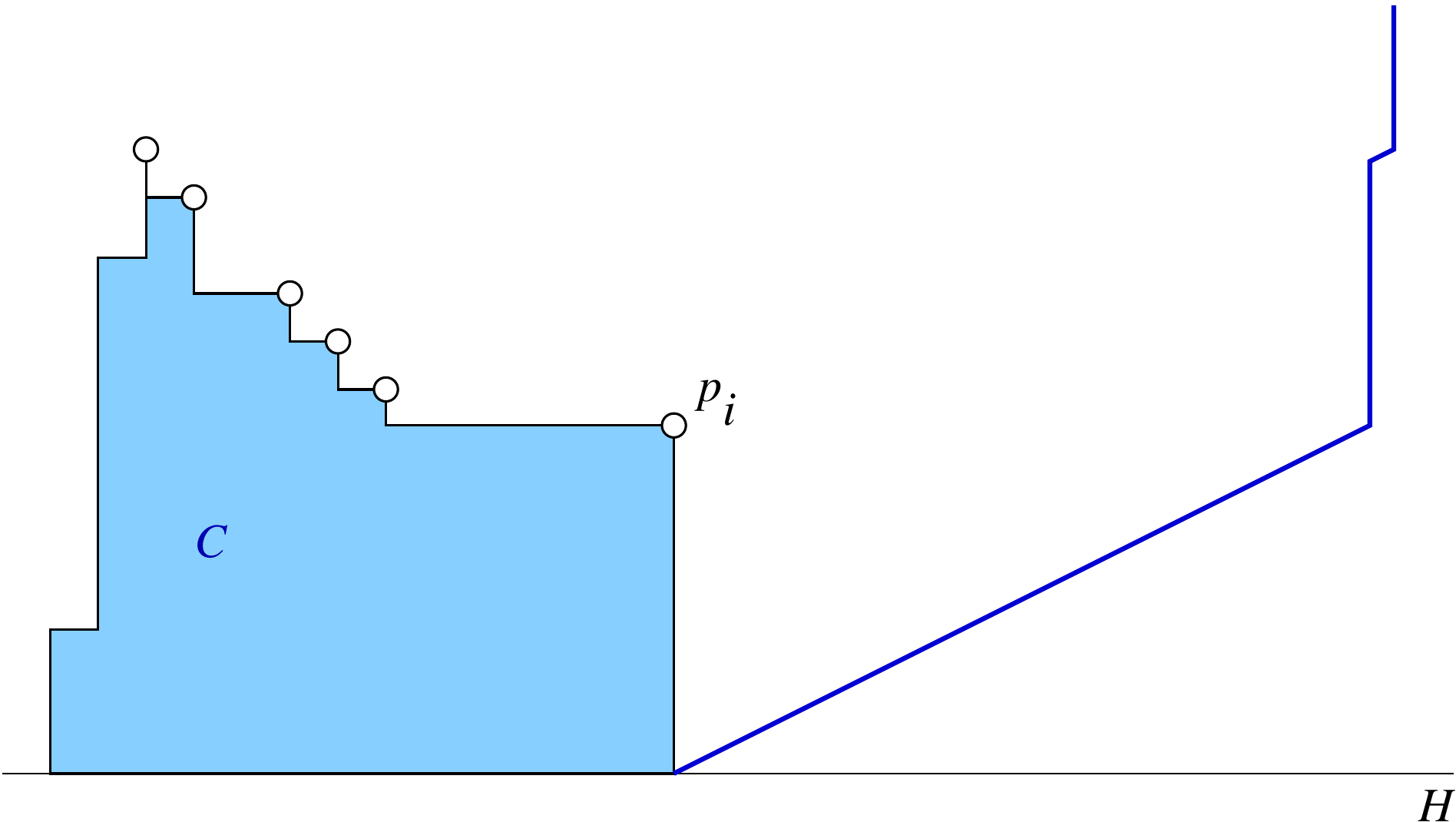}}
\caption{\label{fig:merge}Creating a new cluster.}
\end{figure}

Merging the clusters involves removing the lists of the clusters below $t$ from our set of available lists. This takes $O(k)$ time, where $k$ is the number of clusters merged.
Then we include the walking region of $p_i$ in the list of $\wrb (\C)$. This step might involve deleting a prefix of that list and is illustrated
in Figure~\ref{fig:merge}.

The total time for deleting all segments is $O(n)$, because each segment corresponds to a point and can be deleted only once.
We can use the current number of clusters as a potential function, to show that each merging step
takes $O(1)$ amortized time. Hence the overall complexity of the sweep is $O(n)$, provided that the points are sorted on the $x$-axis beforehand.

The procedure presented thus far groups the point set into clusters. To output the highway hull, we compute the orthogonal convex hull of each cluster.
It remains to determine if at least one pair uses the highway. During the procedure described above, we also maintain at the same time the common intersection of the walking regions of all points. This region is delimited on each side by one sloped segment and by one vertical half-line; we can thus update it in $O(1)$ time per point inserted. At the end of the procedure, we scan the whole set of points a second time and check if every point was contained in that common intersection. If this is the case, no pair uses the highway. 

If at least one pair of points uses $H$, we attach all convex regions to $H$ with vertical segments. If there are points on both sides of $H$, we apply the algorithm to each side, and join both to $H$. 

The complexity is thus $O(n \log n)$. Note that an easy lower bound of $\Omega(n \log n)$ comes from the fact that computing the highway hull is at least as hard as computing the standard $L_{1}$ convex hull, because when no pair of points uses the highway, both problems are equivalent. 
\begin{theorem}
The orthogonal highway hull can be computed in $\Theta(n\log n)$ worst-case time using $\Theta(n)$ space.
\end{theorem}

\section{The Euclidean Highway Hull}
\label{sec:euclidean}

Properties of the Euclidean highway hull are detailed in~\cite{P03,YL07}.
We concentrate on the properties required for our algorithm.

\subsection{Model}
\begin{figure}
\center\includegraphics[scale=1]{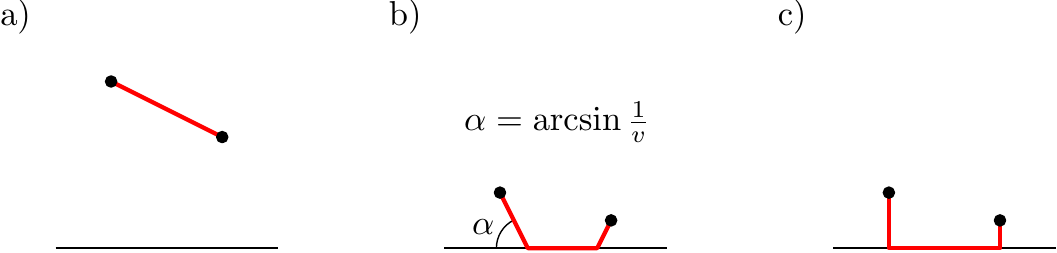}
\caption{\label{fig:euclideansp}Shortest paths in the Euclidean highway model: (a) a path not using $H$, (b) a path using $H$ with bounded speed $v$, (c) a path using $H$ with infinite speed.}
\end{figure}

Without loss of generality, and to simplify the exposition, we will assume that $H$ is on the $x$-axis, all points are above $H$, and that at least one pair uses it.
The shortest path between two points is either the Euclidean line segment or a three-part piecewise linear path (see Figure~\ref{fig:euclideansp}). 
\noindent A key property of shortest paths that use $H$ is that they 
obey Snell's law of refraction, and therefore the angle of incidence of the line segments and the highway is equal to $\alpha = \arcsin (1/v)$.

\begin{figure}
\begin{center}
\includegraphics[scale=1]{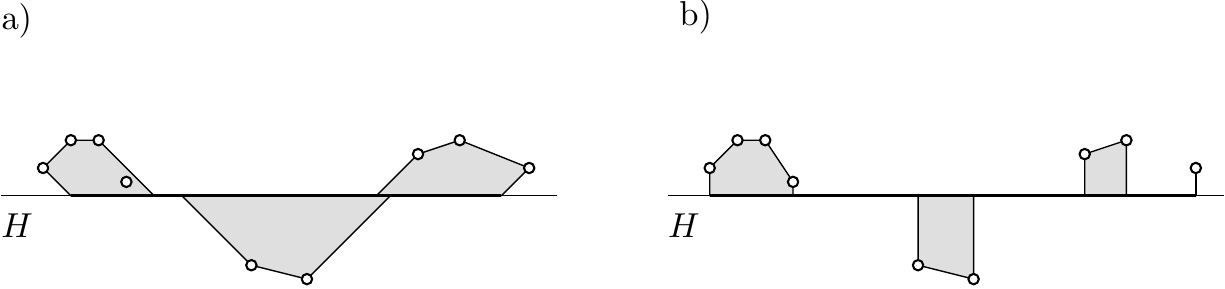}
\end{center}
\caption{\label{fig:L2hull}Illustration of the $L_{2}$ highway hull: (a) finite speed $v$, (b) infinite speed.}
\end{figure}

The definition of the Euclidean highway hull $\HH_2 (S)$ of a point set $S$ is similar to that of $\HH_1(S)$: it is the minimal (under inclusion) set $R\supseteq S$ containing all shortest paths between pairs of points of $R$ (i.e.,~the closure under the operation of including shortest paths). Figure~\ref{fig:L2hull} illustrates $\HH_2 (S)$. 

\subsection{Observations}
Many properties of the orthogonal highway hull are still true in the Euclidean case: it has no holes, and if at least one pair uses $H$ then $\HH_2 (S)$ contains the ``slanted'' projections of all $n$ points onto $H$, i.e.,~projections in one of the directions given by Snell's law of refraction.

\begin{figure}
\begin{center}
\includegraphics[scale=0.7]{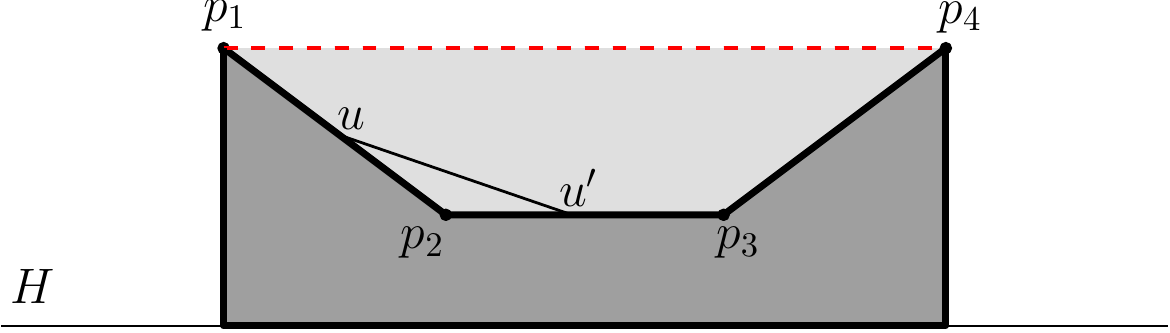}
\end{center}
\caption{\label{fig:closed}The Euclidean highway hull is not always a closed set. Here, $v$ is infinite on $H$.}
\end{figure}

An interesting new property is that $\HH_2(S)$ is not always a closed set. 
Consider the set of four points $(p_{1},p_{2},p_{3},p_{4})$ in Figure~\ref{fig:closed}. $H$ is not used between $p_{1}$ and $p_{2}$. The same is true for $p_{2},p_{3}$ and for $p_{3},p_{4}$. Thus the highway hull must contain the polygonal path $p_{1},p_{2},p_{3},p_{4}$. However $H$ is used from $p_{1}$ to $p_{4}$. Since at least one pair uses $H$ and the hull has no holes, the region between the polygonal path and $H$ (shown in dark gray in the figure) is contained in the hull.

Since the highway hull is the closure under the operation of including shortest paths, the construction of the hull in our example is incomplete. Consider any reflex vertex on the boundary of the convex hull, such as $p_{2}$. Take two points $u$ and $u'$ on the boundary of the hull, arbitrarily close to $p_{2}$, with $u$ to the left of $p_{2}$ and $u'$ to its right. 
If $u$ and $u'$ are close enough, $H$ is not used to go from one to the other. Thus the segment $uu'$ belongs to the highway hull, as does the region between $uu'$ and $H$. This proves that the highway hull has no reflex vertices (except at the junction of a cluster with $H$).

Our preceding arguments imply that the rectangle formed by $p_1,p'_1,p_4,p'_4$ is contained in the highway hull. However the segment $p_{1}p_{4}$ is \emph{not} in the highway hull. As mentioned, $H$ is used between $p_{1}$ and $p_{4}$. There is no pair of points $u,u'$ within the hull that have a shortest path intersecting the segment $p_{1}p_{4}$. Thus the closure operation will never include points on the open segment $p_{1}p_{4}$. 
Every other point in the region bounded by $p_{1}p_{4}$ and the polygonal path $p_{1},p_{2},p_{3},p_{4}$ (i.e.,~the region in light gray in the figure) is added by closure.

In what follows, we will ignore this issue about open and closed edges, i.e., we will propose an algorithm which finds the correct hull, except that some of its boundary edges should be open. At the end of the procedure, we will identify these open edges.\medskip

When the Euclidean metric is used, the walking region $W(p)$ of a point $p$ is delimited by parabolic segments (see~\cite{P03}, Proposition~2.6.1). In the specific case of $v=+\infty$, the right boundary of the walking region $\wrb (p)$ of a point is a single parabola tangent to $H$. To simplify the exposition, we focus on solving the problem when the speed is infinite. A description of the algorithm for finite speed is given at the end of this section.

\begin{figure}[h!]
\begin{center}
\subfigure[\label{fig:parabola}The walking region of a new point $p_4$ happens to intersect an edge of cluster $\mathcal C$ without containing any of its vertices.]{\includegraphics[clip=true, trim= 0pt 0pt 0pt 5cm, scale=.4]{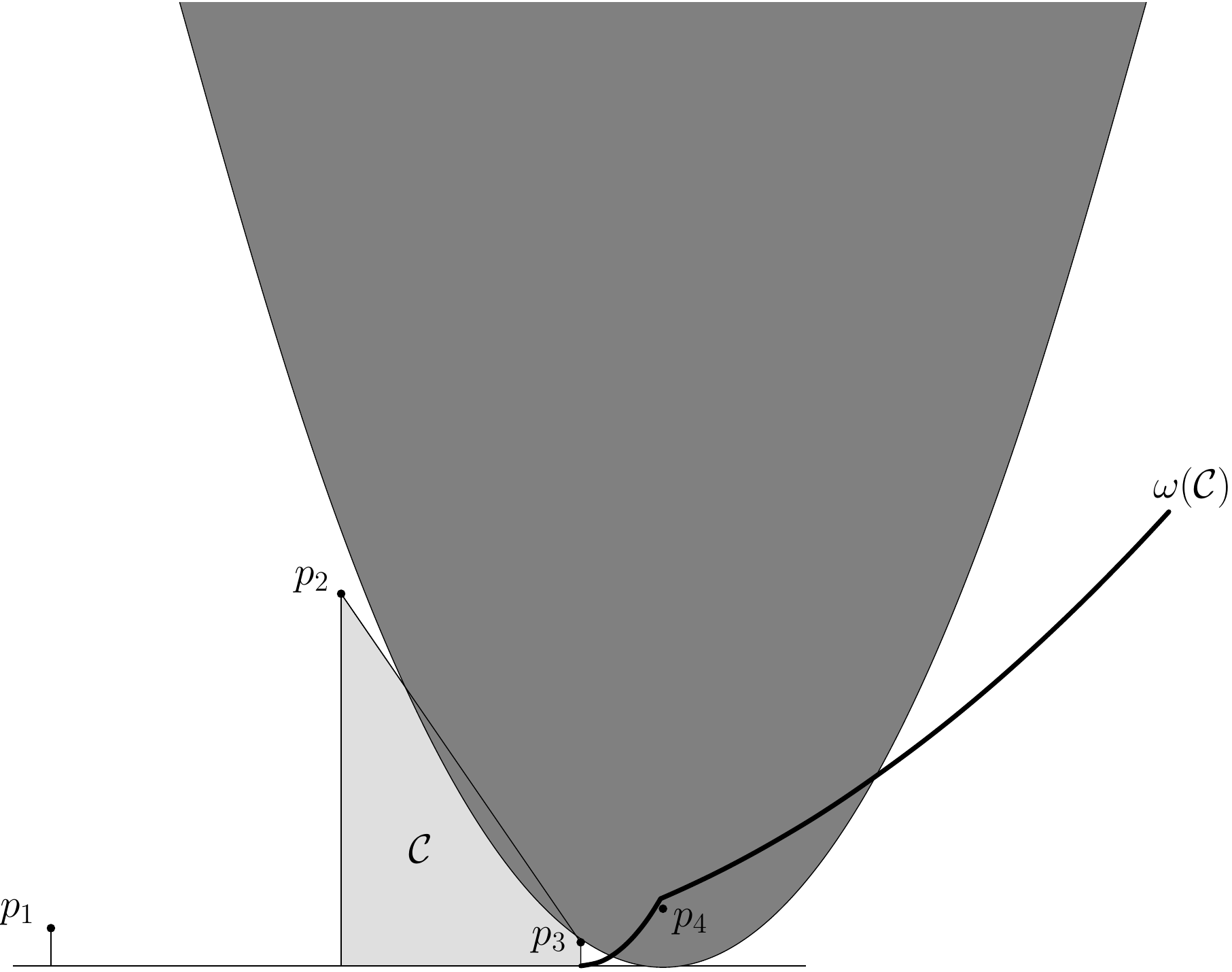}}
\hspace{1cm}
\subfigure[\label{fig:newsegment}The interior of a new segment $p_2p_3$ intersects the walking region of $p_1$.]{\includegraphics[clip=true, trim= 0pt 0pt 0pt 2cm, scale=.5]{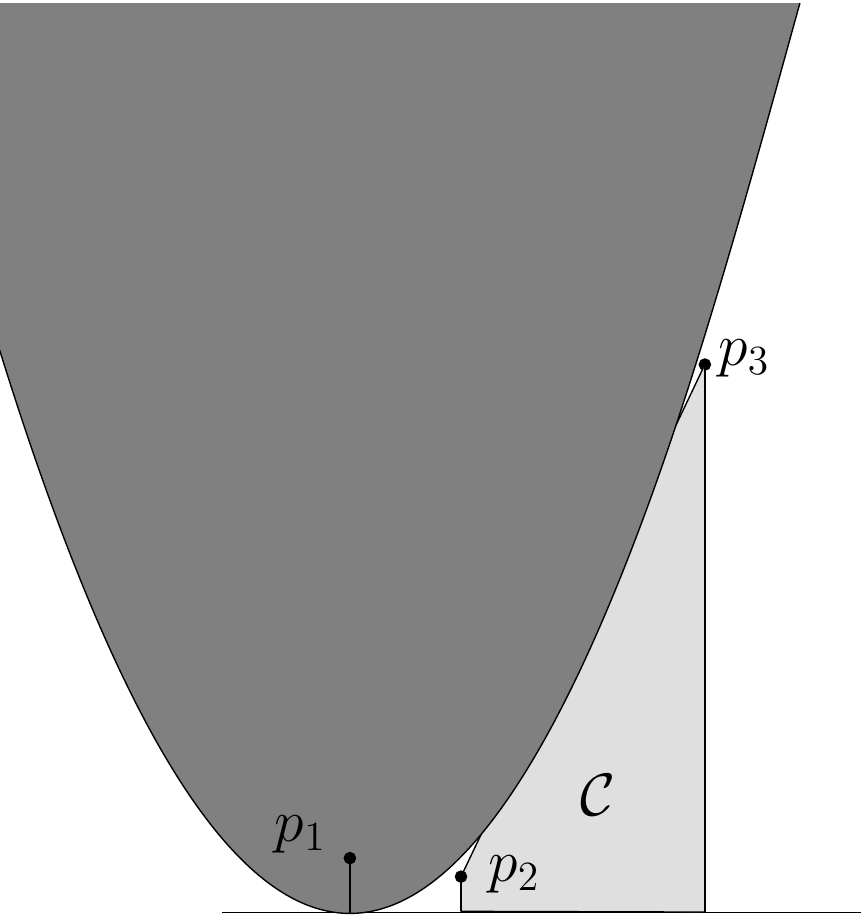}}
\end{center}
\caption{\label{fig:examples}Two cases in which the algorithm in~\cite{YL07} fails.}
\end{figure}

In the orthogonal case, Observation~\ref{lem:vtx} allowed us to restrict our attention to the extreme points of the highway hull
when partitioning the points in clusters. However, Figure~\ref{fig:examples} exhibits two situations that 
preclude
the application of the previous algorithm to the Euclidean case. These situations are not
 taken into account in the algorithm of Yu and Lee~\cite{YL07}.

In the first situation, shown in Figure~\ref{fig:parabola}, a new point $p_4$ is outside the right boundary 
$\wrb(\C)$
of the walking regions
of vertices $p_2$ and $p_3$. Its walking region $\wrb (p_4)$ intersects segment $p_2 p_3$, and so $p_4$ should be
merged into the cluster of its two predecessors. This is not detected unless we correctly handle the walking region of the entire
segment $p_2p_3$.

In the symmetric situation of Figure~\ref{fig:newsegment}, the newly considered point $p_3$ has just been merged 
into a cluster with $p_2$. Both $p_2$ and $p_3$ are outside $\wrb (p_1)$. 

\noindent Since a cluster is a convex set, we must include all points of edge $p_2p_3$ in the new cluster. 
However, at least one such point $x$ intersects $\wrb (p_1)$,
and so the shortest path between $p_1$ and $x$ is a single line segment.
Now this segment $p_1 x$ creates new points in the cluster,
which must also be considered in the recursion that computes the closure to obtain the new hull. 

\begin{figure}
\begin{center}
\includegraphics[scale=.7,clip=true,trim = 0pt 0pt 0pt 8cm]{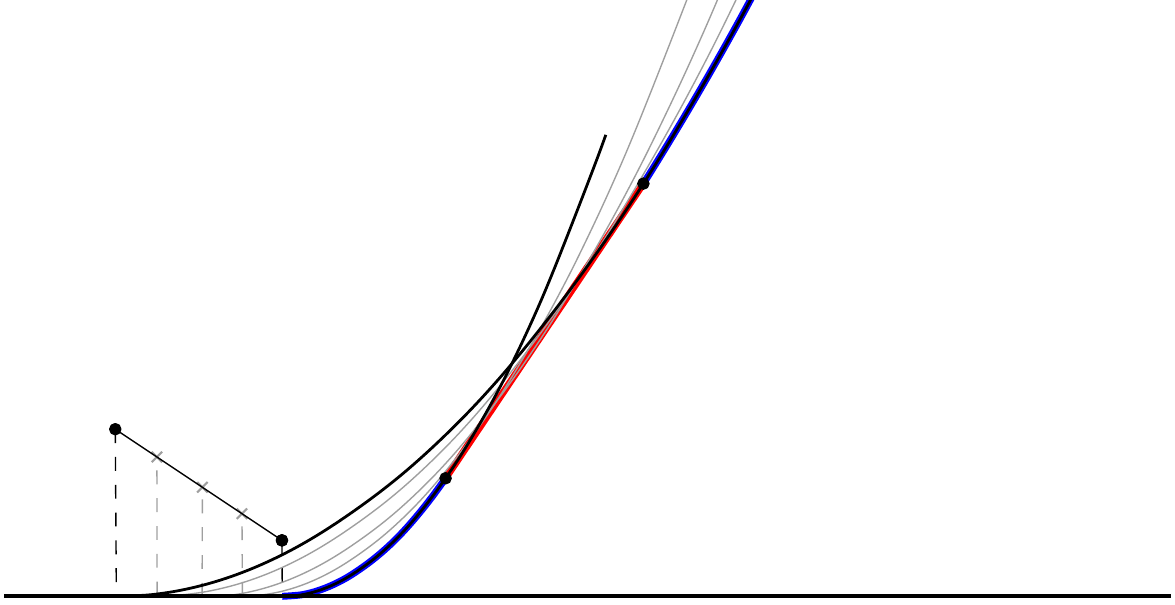}
\end{center}
\caption{\label{fig:psp}The boundary of the walking region of an edge.}
\end{figure}

The first situation can be handled by considering the boundaries of the walking regions of the edges of the hull, not only of the vertices.
For $v=+\infty$, the walking region of a segment is the convex hull of the walking regions of its endpoints. The right boundary is a
three-part convex curve, consisting of two parabolic arcs joined by a segment that is tangent to both 
(see Figure~\ref{fig:psp}). The curve is denoted by $\wrb (ab)$, where $a$ and $b$ are the points of tangency. A formal description of the walking region boundary structure is given in~\cite[Lemma 2.1.9 to 2.1.12]{P03}.

Other useful properties of the right boundary of a cluster's walking region include~\cite{YL07}:
\begin{itemize}
\item the right boundary is $x$- and $y$-monotone.
\item the indexing of parabolic segments along the right boundary is inverted with respect to their corresponding points. In other words, the leftmost parabolic segment corresponds to the walking region of the cluster's rightmost point.
\item Lemma~\ref{lem:inbetween} is still valid in the Euclidean case: if $x$ is a convex vertex of $\mathcal C$ whose
walking region contains a point $p$, then all the clusters between $\mathcal C$ and $p$ are below the shortest path from $p$ to $x$.
\end{itemize}

\subsection{Algorithm: Preliminary Lemmas}
As in the orthogonal case, the algorithm is composed of two main loops. The outer loop considers points $p_i$ in sorted order of their projections on $H$. The inner loop identifies the cluster 
with which $p_i$ should be merged. In order to simplify the exposition, the algorithm for computing $\HH_2(S)$ will be based on the algorithm for $\HH_1(S)$ (similar to that of Yu and Lee~\cite{YL07}). We will use an additional data structure to
handle the problematic situation of Figure~\ref{fig:newsegment}. 
We first explain how to apply the previous algorithm.\\

Each cluster $\mathcal C$ is associated with a boundary $\wrb (\C)$. This boundary is stored in a list, each element of which encodes the boundary of the walking region of an edge of $\mathcal C$. Observation~\ref{lem:vtx} also holds in the Euclidean case (a similar observation was made by Yu and Lee~\cite{YL07}). This implies that $\wrb (\C)$ is defined
solely by the walking regions of the {\em negatively sloped} segments of $\mathcal C$.

\begin{figure}
\begin{center}
\includegraphics[scale=.7,clip=true,trim = 0pt 0pt 5cm 13.3cm]{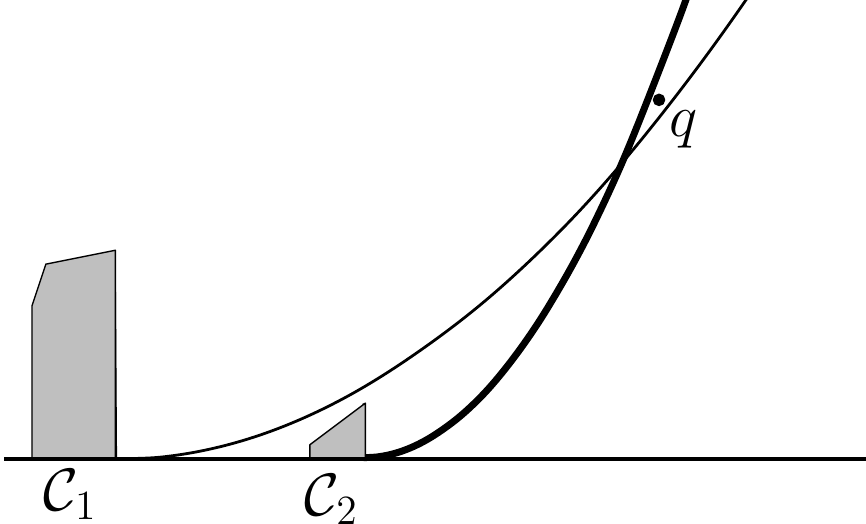}
\end{center}
\caption{\label{fig:nolem3}Although $q$ is below the right boundary of the walking region of the rightmost cluster ($\C_{2}$), it is above the right boundary of the walking region of $\C_{1}$. This implies that $\C_{1},\C_{2},$ and $q$ will all be merged into a single cluster.}
\end{figure}

The main issue, however, is that Lemma~\ref{lem:order} 
(vertical ray crossing order)
does not hold in the Euclidean setting. In other words, we cannot just check if a new point belongs to the rightmost cluster to determine if a new cluster must be created (see Figure~\ref{fig:nolem3}).

\begin{lemma}\label{heavenandhell}
The right boundaries $\wrb (\C_i)$ and $\wrb (\C_j)$ of two walking regions
of two clusters 
intersect at most once.
\end{lemma}

\begin{proof}
For the purpose of contradiction, 
suppose that $\wrb (\C_i)$ and $\wrb (\C_j)$ intersect at least twice, at $i_{1}$ and $i_{2}$. Let $a \prec_{x} b$ ($a \prec_{y} b$ respectively) denote that the $x$-coordinate ($y$-coordinate respectively) of $a$ is less than that of $b$. Without loss of generality, let $i<j$ and $i_{1} \prec_{x} i_{2}$. 

We can identify four points $p_{1},p_{2},p_{3},p_{4}$ such that: 
$$p_{1},p_{2} \in \C_{i}$$
$$p_{3},p_{4} \in \C_{j}$$
$$i_{1},i_{2} \in \wrb(\C_{i} \cup \C_{j}) $$
$$\wrb( p_{1})\cap \wrb (p_{3}) = i_{1}$$
$$\wrb( p_{2})\cap \wrb (p_{4}) = i_{2}$$
The walking regions of these four points pass through $i_1$ and $i_2$ (see Figure~\ref{fig:heavenandhell}). 
Note that $p_{1}$ and $p_{2}$ need not be distinct (same for $p_{3}$, $p_{4}$).
Since clusters are disjoint, we know that:
$p_{1}\prec_{x} p_{3}$, $p_{2}\prec_{x} p_{3}$, $p_{1}\prec_{x} p_{4}$, and
$p_{2}\prec_{x} p_{4}$.
Since walking region boundaries are 
$x$-monotone, 
and boundary segments have an inverted ordering with respect to the points/segments that created them, 
we know that $p_{2}\preceq_{x} p_{1}$, and $p_{4}\preceq_{x} p_{3}$.
Thus we obtain the following order along the $x$-axis: 
$$p_{2} \preceq_{x} p_{1}\prec_{x} p_{4}\preceq_{x} p_{3}$$
As $i_{1}$ and $i_{2}$ are distinct, $p_{1} \preceq_{y} p_{2}$. Otherwise $\wrb( p_{2})$ would be to the left of $\wrb( p_{1})$ and thus left of $\wrb (\C_i)$. 
Similarly, $p_{3} \preceq_{y} p_{4}$.
As $\wrb( p_{1})$ intersects $\wrb (p_{3})$, and $p_{1}\prec_{x} p_{3}$, we deduce that $p_{3} \prec_{y} p_{1}$. Similarly, $p_{4} \prec_{y} p_{2}$.

\begin{figure}
\begin{center}
\includegraphics[scale=.5]{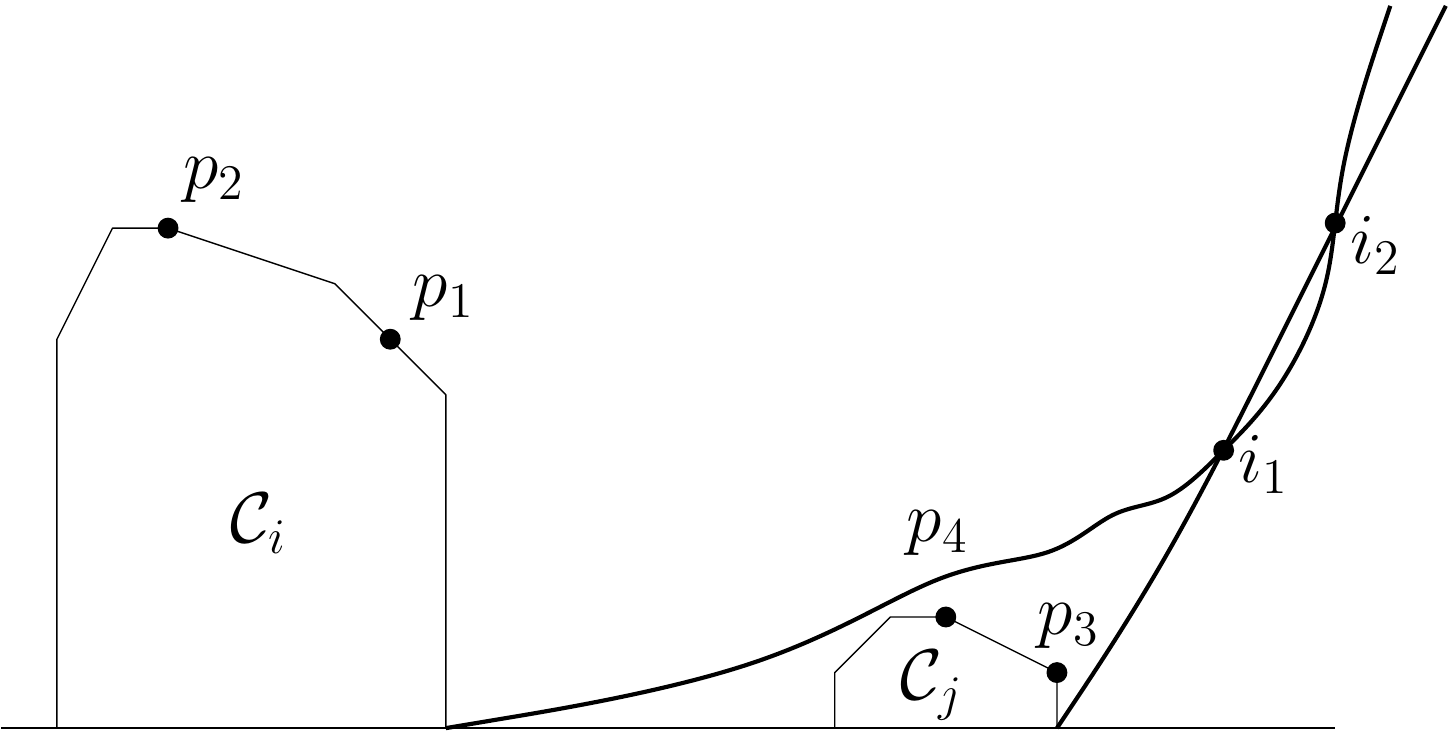}
\end{center}
\caption{\label{fig:heavenandhell}
Lemma~\ref{heavenandhell} establishes that
the right boundaries of the walking regions of two clusters intersect at most once.
}
\end{figure}

Notice that $\wrb( p_{1})$ cannot be to the left of $\wrb (p_{4})$; this would imply that the intersection point $i_{1}$ between $\wrb( p_{1})$ and $\wrb( p_{3})$ is to the left of $\wrb (p_{4})$, and thus to the left of $\wrb (\C_j)$, contradicting the fact that $i_{1}\in \wrb (\C_j)$. We conclude that $p_{4} \prec_{y} p_{1}$, and thus 
$$p_3\preceq_{y}p_{4}\prec_{y}p_{1}\preceq_{y}p_{2}$$

\begin{figure}
\begin{center}
\includegraphics[scale=.5]{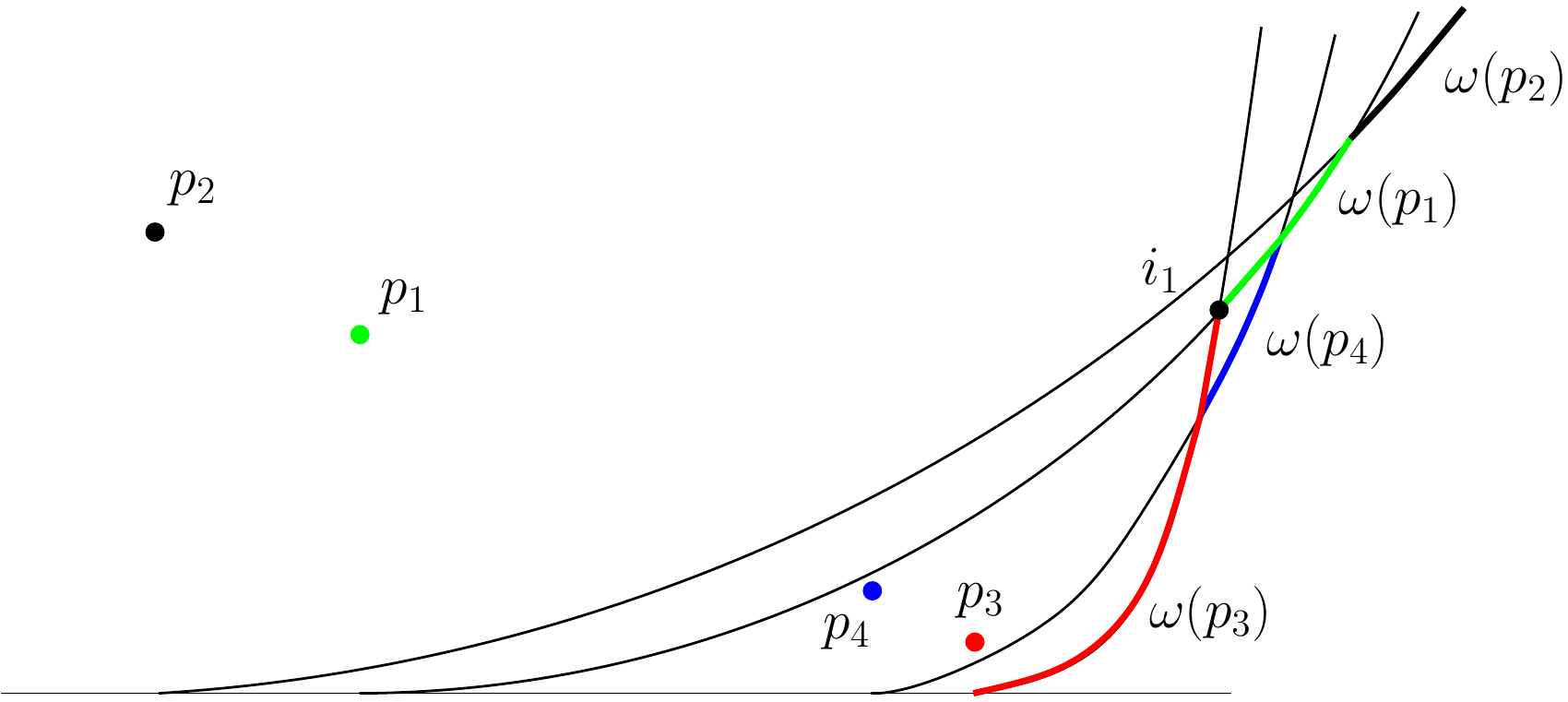}
\end{center}
\caption{\label{fig:hah2}The right boundary of the walking region of the four points ${p_{1},p_{2},p_{3},p_{4}}$.}
\end{figure}

With these relationships established, consider
the walking region of all four points, $\wrb (\{p_{1},p_{2},p_{3},p_{4}\})$. 
It is composed of four parabolic segments, 
ordered inversely with respect to the points (see Figure~\ref{fig:hah2}).
This implies that the intersection $i_{1}$ between $\wrb(p_{3})$ and $\wrb(p_{1})$ occurs strictly to the left of $\wrb (\{p_{1},p_{2},p_{3},p_{4}\})$, since $\wrb(p_{3})$ and $\wrb(p_{1})$ do not appear consecutively on $\wrb (\{p_{1},p_{2},p_{3},p_{4}\})$. Thus $i_{1}$ is to the left of $\wrb (\C_{i} \cup \C_{j})$,
which contradicts its definition of existing on that boundary.
\qed\end{proof}

\noindent We now have all the tools needed to prove an analogue of Lemma~\ref{lem:order} for the Euclidean case. 
For a point $p$, let the set $R(p)$ be defined as follows:
$$R(p)=\left\{\wrb(\C_{i})| \forall i'<i, p\prec_{x} \wrb(\C_{i})\cap\wrb(\C_{i'})\right\}$$
\noindent $R(p)$ is the set of all right boundaries whose intersections with previous boundaries are to the right of $p$. 
\begin{lemma}\label{lem:newlem3}
The boundaries in $R(p)$ cross a vertical ray from $p$ in a 
right-to-left order of the corresponding clusters.
\end{lemma}

\begin{proof}
The proof is by induction. If there is only one cluster, and thus one boundary, our claim is trivially true. Assume it is true up to the construction of cluster $\C_{i}$. We add a cluster $\C_{j}$ to the right of $\C_{i}$ (see Figure~\ref{fig:order}). 
\begin{figure}
\begin{center}
\includegraphics[scale=.8,clip=true,trim = 0pt 0pt 7cm 20cm]{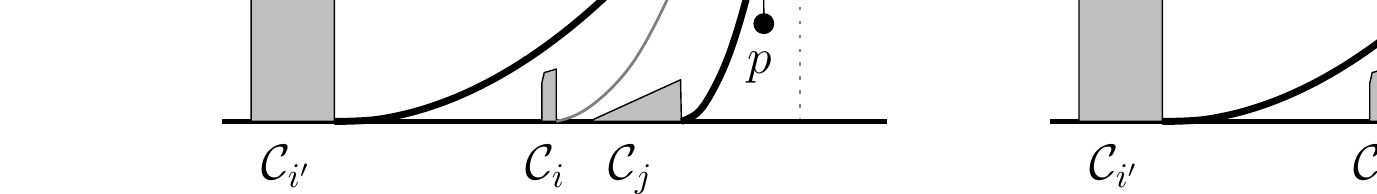}
\end{center}
\caption{\label{fig:order} (a) $p\prec_{x} t$ and $R(p)=\{\wrb(\C_{i'}),\wrb(\C_{j})\}$. (b) $t\prec_{x} p$ and $R(p)=\{\wrb(\C_{i'})\}$. $R(p)$ is illustrated in bold.}
\end{figure}

\noindent By Lemma~\ref{heavenandhell}, $\wrb(\C_{j})$ intersects the boundary of each of the previous clusters at most once. Let $t$ be the leftmost intersection of $\wrb(\C_{j})$, and let $\C_{i'}$ be the cluster whose boundary intersects that of $\C_{j}$ at $t$. $\C_{j}$ must be below $\wrb(\C_{i'})$, otherwise $\C_{i'}$ and $\C_{j}$ would not be disjoint. Thus $\wrb(\C_{j})$ is below $\wrb(\C_{i'})$ to the left of $t$. By the same reasoning, $\wrb(\C_{j})$ is below every other boundary to the left of $t$. Now we have two cases :

Case 1: $p \prec_{x} t$ (see Figure~\ref{fig:order}(a)). The vertical ray emanating from $p$ will first intersect the boundary of the rightmost cluster $\wrb(\C_{j})$. It will then intersect the boundaries of previous clusters. By induction, those in $R(p)$ will be intersected in right-to-left order, and we deduce the claim.

Case 2: $t\preceq_{x}p$ (see Figure~\ref{fig:order}(b)). Then $\wrb(\C_{j})$ is not in $R(p)$, and thus by induction the claim is true, as the new cluster can be ignored.
\qed\end{proof}

We modify the algorithm used for $L_{1}$ as follows: for every new cluster, we compute the first intersection of the right boundary of its walking region with the right boundaries of walking regions of all previous clusters. The boundary $\wrb (\C_i)$ for a cluster $\C_i$ can be represented by a linked list $L_i$ of parabolic segments and line segments. As before, we maintain a set $\mathcal L$ of non-empty lists, each associated with a cluster, representing boundaries that have not yet been entirely scanned. We do not store the portions of boundaries to the right of their first intersection point with the boundaries of previous clusters.

By Lemma~\ref{lem:newlem3}, the segments in all cluster lists are crossed by any vertical ray in right-to-left order. Excluding the right boundaries after the intersection point is not a problem: if a point $p$ is above the right boundary $\wrb (\C_i)$ of some cluster $\C_i$ after its intersection with $\wrb (\C_{i'})$, where $i'<i$, then the leftmost cluster containing $p$ is not $\C_i$, but $\C_{i'}$. Thus, from that point onwards, the boundary of $\C_i$ can be discarded.

\subsection{Algorithm Outline}

Let $\C_j$ be the rightmost cluster. The algorithm starts by scanning the 
list $L_j$ from left to right. At each step, we consider the relative position of $p_i$ and the current 
segment $s$ in the list, and we apply the same four-step algorithm as for the orthogonal case. 
We end up with a point $x$ corresponding to the segment in the leftmost cluster 
which must be merged into $\C_j$
by the addition of edge $x p_{i}$. 

\begin{figure}
\begin{center}
\subfigure[The new segment has negative slope.]{
\includegraphics[angle=-90,scale=.5]{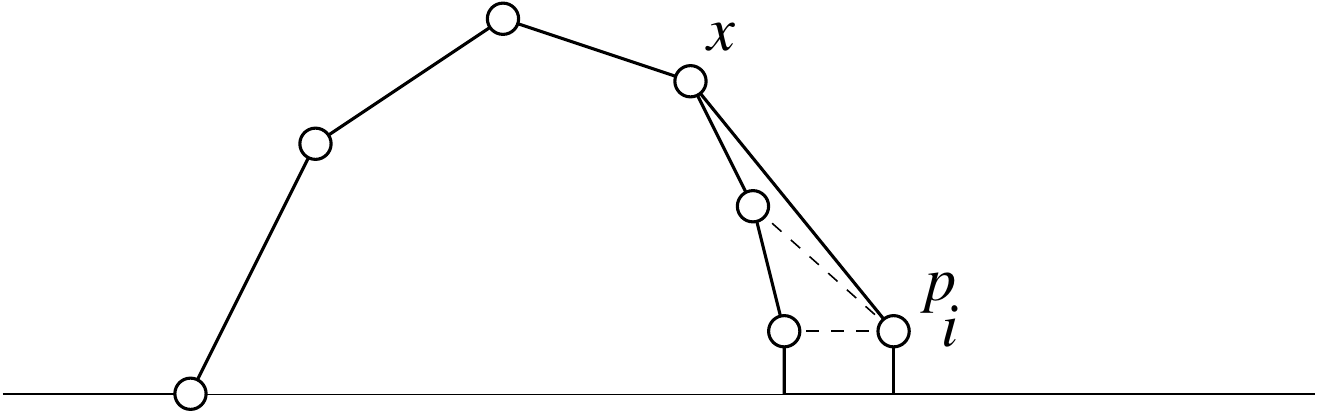}}
\hspace{.5cm}
\subfigure[The new segment has positive slope.]{\includegraphics[angle=-90,scale=.5]{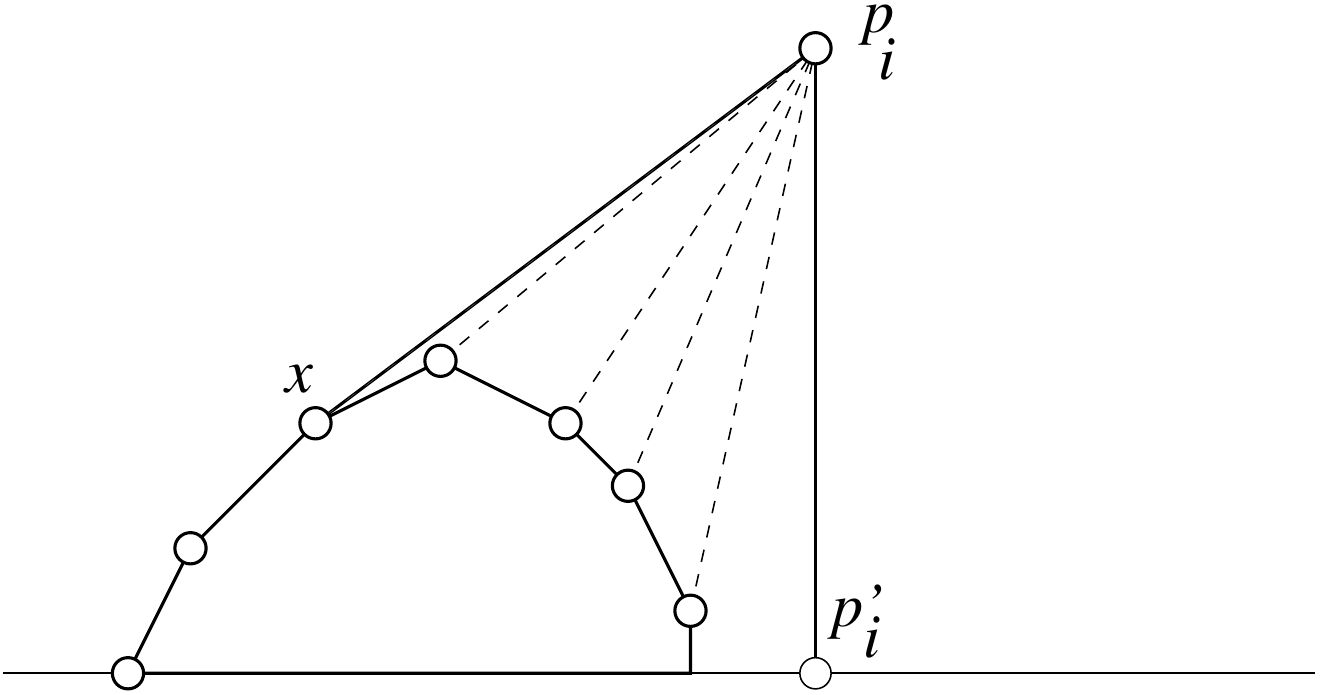}}
\end{center}
\caption{\label{fig:slopenewseg}The two cases occurring as we insert a new segment.}
\end{figure}

Since clusters are convex, if the edge we want to insert creates a reflex angle, we scan the boundary of the cluster up to the point where we get a new convex cluster (see the dotted lines in Figure~\ref{fig:slopenewseg}). 
We distinguish two cases, depending on the slope of $x p_i$.
If the slope is negative, then it will never trigger the 
problematic situation of Figure~\ref{fig:newsegment}; a right parabolic walking region boundary can only cut into a positively sloped segment.
The only remaining task is to update the lists 
$\mathcal L$
by taking $\wrb (x p_i)$ into account.

On the other hand, if $x p_i$ has a positive slope, then we check if it intersects the walking region boundary of another cluster, as in Figure~\ref{fig:newsegment}.
Note that in that case, the right boundary of the walking region of the new cluster 
consists solely of $\wrb (p'_ip_i)=\wrb (p_i)$. This is because the new point will be above and to the right of all previous points. Thus by the analogue of Observation~\ref{lem:vtx}, the boundary will be defined by the upper-rightmost point (see~\cite{YL07}). 

Also, note that adding one point results in the addition of at most one edge, since Lemma~\ref{lem:inbetween} is valid in $L_{2}$. This implies that we only need to add the edge between the new point and the leftmost cluster; all intermediate clusters will be below the new edge, and will disappear. 

In order to efficiently answer the segment intersection queries
between $x p_i$ and the segments in ${\mathcal L}$,
we maintain a data structure
that will store a representation of $\wrb = \wrb (\bigcup_i \C_i)$ (i.e.,~the right boundary of the union of all walking regions). By Observation~\ref{lem:vtx}, this boundary only depends on the negatively sloped segments of the hull. 
The data structure must be able to answer segment intersection queries to maintain the representation of $\wrb$ under the following two operations: {\em rollback}, which removes the rightmost negatively sloped segment, and {\em insertion}, which adds a new segment to the right of the current hull.

Rollbacks are performed every time an edge is deleted during a merge operation. If the cost of a rollback is $T(n)$ in the worst case, then it costs only $O(nT(n))$ over the whole algorithm, since each time an edge is deleted, either a point or its projection on $H$ disappears. An insertion is performed when the new segment $x p_i$ has a negative slope, in which case we insert $xp_i$, or when $xp_i$ has positive slope but empty intersection with $\wrb$, in which case we insert the segment $p'_i p_i$. We denote by $Q(n)$ and $I(n)$ the costs of a segment intersection query and an insertion, respectively.

When $xp_i$ intersects $\wrb$ (i.e.,~when the intersection query returns a positive result), we identify the cluster causing this intersection, and iterate with a new segment $x'p_i$. This involves deleting more edges in the boundary lists, and performing the corresponding rollbacks in the data structure for $\wrb$.

\subsection{Segment Intersection Queries}

Segment intersection queries can be performed using ray shooting. Instead of considering a segment $xp_i$ itself, we consider the ray emanating from $p_i$ and containing $xp_i$. If the ray intersects the boundary $\wrb$, we can quickly check whether the intersection points belongs to $xp_i$. Note that all queries consist of rays that are directed down and to the left. We can also assume that $p_i$ is below $\wrb$. 

General dynamic data structures exist for ray shooting (see for instance 
~\cite{goodrich1997drs}). However, they are designed for planar subdivisions with line segments only. We proceed to describe a static data structure for answering ray shooting queries in our setting.

\subsubsection{A ray shooting data structure.}

The input data is a sequence of parabolic arcs and segments that compose the boundary $\wrb$. We decompose this sequence recursively and store the decomposition in a binary tree $T$ of height $O(\log n)$. Each node $t\in T$ corresponds to a portion $\wrb (t)$ of the curve $\wrb$, and is associated with the lower convex hull $CH(t)$ of $\wrb (t)$. We use a suitable data structure for $CH(t)$ that can answer ray shooting queries in $O(\log n)$ time (see for instance~\cite{rayshoot}). The root of $T$ is denoted by $r(T)$, and the left and right nodes of $t$ are denoted by $\leftn (t)$ and $\rightn (t)$. The left and right nodes correspond to the top and bottom part of $\wrb$, respectively.

Ray shooting queries on $\wrb$ are answered by a simple traversal of $T$.
We first suppose that $p_i\not\in CH(r(T))$, meaning the origin $p_i$ of the ray is not contained in the convex hull of $\wrb$. 
We iterate the following operations starting from $r(T)$, until either we hit the curve or we conclude that the intersection is empty. Let $t$ be the current node. If the ray intersects $CH(\leftn (t))$, we iterate with $\leftn (t)$, and we do not need to iterate with $\rightn (t)$, because the intersection with $\leftn (t)$ will happen before that with $\rightn (t)$. Otherwise if it intersects $CH(\rightn (t))$, we iterate with $\rightn (t)$. If neither intersection occurs, we are done. Since $T$ has height $O(\log n)$ and each intersection test requires $O(\log n)$ time, the whole algorithm takes $O(\log^2 n)$ time.

If $p_i\in CH(t)$ for some node $t\in T$, we cannot conclude that the ray intersects $\wrb (t)$. Hence when $p_i\in CH(r(T))$ we first identify $O(\log n)$ subtrees of $T$ such that: {\it (i)} $p_i$ does not belong to any of the corresponding convex hulls, and {\it (ii)} if the ray intersects $\wrb$, then it intersects one of the convex hulls. We choose these subtrees to correspond to convex hulls entirely contained in the halfplane below the horizontal line through $p_i$. There are $O(\log n)$ such maximal subtrees, the roots of which can be identified in $O(\log n)$ time. For each root $r$, from left to right in the tree (hence from top to bottom on $\wrb(t)$), we check whether the ray intersects $CH(r)$. For the first intersection found, we run the previous algorithm on the corresponding subtree. The remaining subtrees need not be examined since we know that the ray will first intersect a portion of the curve contained in that subtree. Checking intersections takes $O(\log n)$ time, and since there are $O(\log n)$ subtrees, we spend $O(\log^2 n)$ time before finding an intersection. Running the previous algorithm in the subtree takes $O(\log^2 n)$ as well. Thus we have $Q(n) = O(\log^2 n)$ in the worst case.

\subsubsection{Insertion and rollback.}

 Insertions in $T$ occur when the walking region of a new segment $xp_i$ has to be taken into account in the boundary $\wrb$. We must remove subtrees of $T$ corresponding to the portion of $\wrb$ that is strictly contained in the new walking region, and update the convex hull $CH(t)$ of each node $t$ on the path from the root to the new node. We show how to achieve this in $O(\log^2 n)$ worst-case time while keeping $T$ balanced. 
 
\begin{figure}
\begin{center}
\includegraphics[angle=-90,scale=.4]{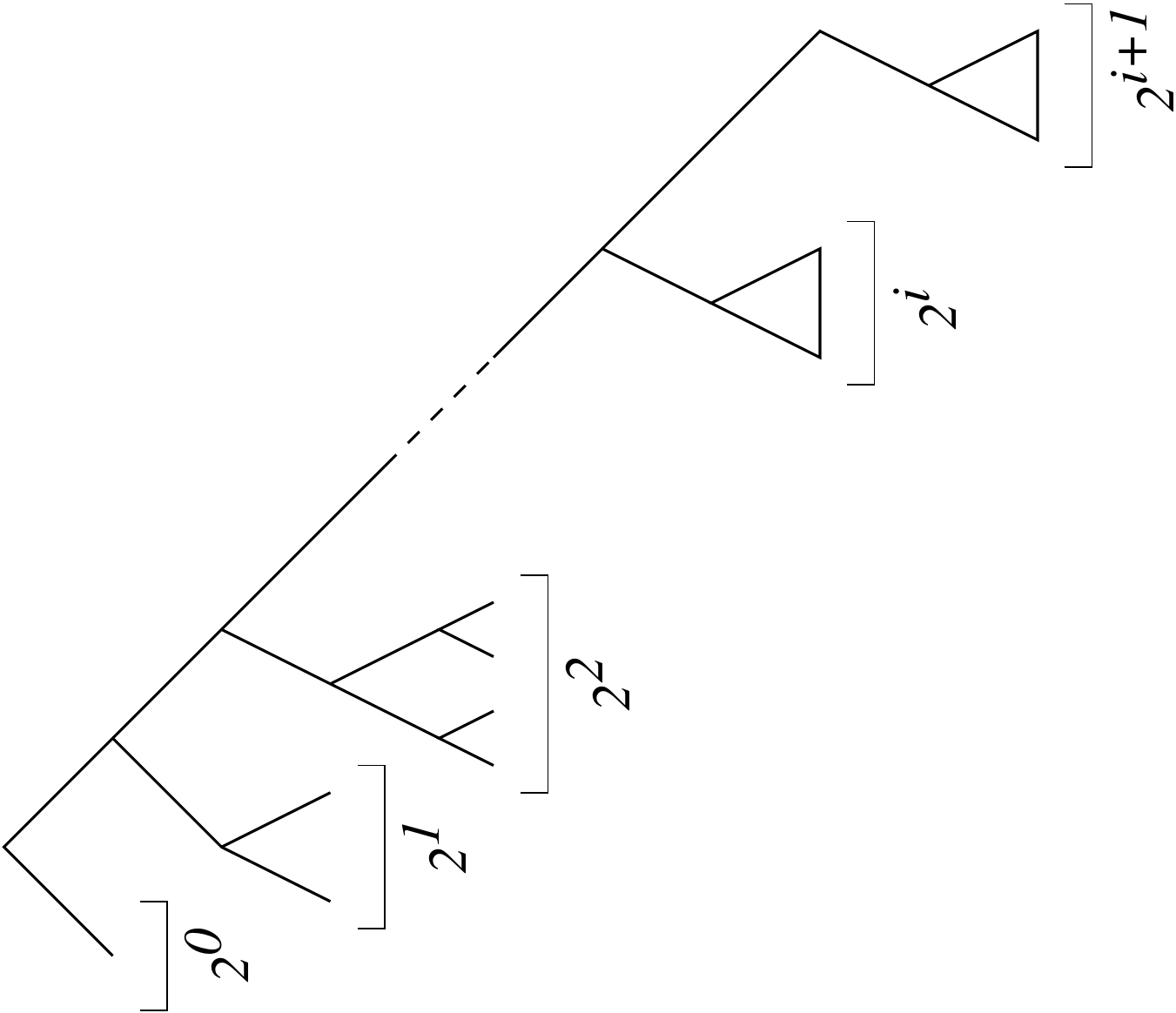}
\end{center}
\caption{\label{fig:exptree}The structure of the tree $T$.}
\end{figure}

Instead of a dynamic balanced tree (such as red-black trees~\cite[chapter 13]{goodrich1997drs,CLRS00}), we use an exponential binary tree, as described in Figure~\ref{fig:exptree}.
The tree is composed of a backbone of right children. The $(i+1)$th left subtree on the backbone is a complete binary tree with $2^i$ leaves, except for the last subtree, the last level of which may be incomplete. It is easy to check that the height of $T$ is $O(\log n)$.

\begin{figure}
\begin{center}
\subfigure[\label{fig:boundaryCH}The boundary $\wrb$, with the convex hulls $CH(r(T))$, $CH(\leftn (r(T)))$, and $CH(\rightn (r(T)))$.]{\includegraphics[angle=-90,scale=.3]{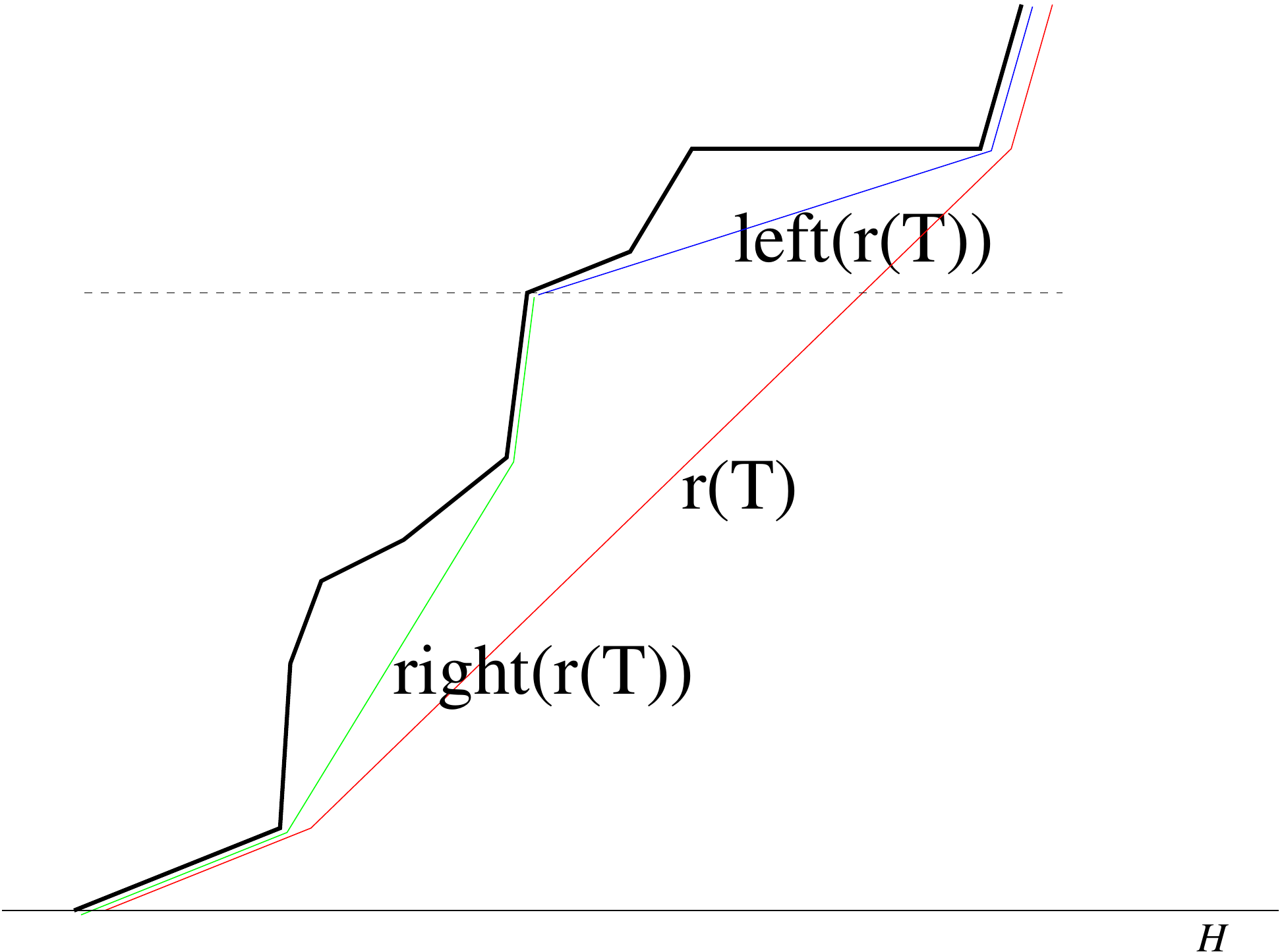}}
\hspace{2cm}
\subfigure[\label{fig:boundaryCHupdate}Insertion: the right subtree rooted at $CH(\rightn (r(T)))$ is deleted.]{\includegraphics[angle=-90,scale=.3]{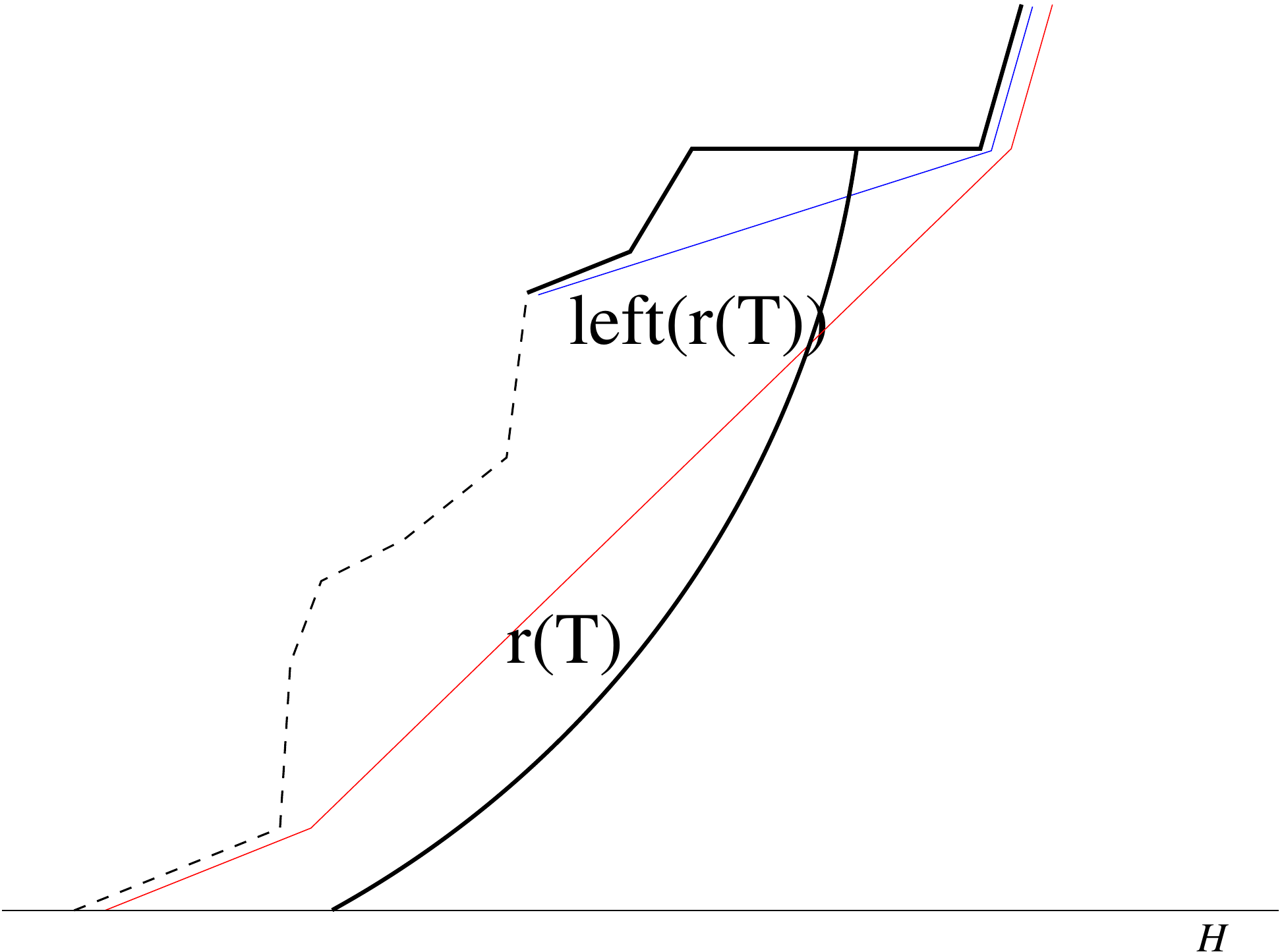}}
\end{center}
\caption{Representation of the boundary $\wrb$.}
\end{figure}

The insertion algorithm uses lazy deletion for parts of the curve $\wrb$ masked by $\wrb (xp_i)$. We iterate the following operations starting from $r(T)$. Let $t$ be the current node. We first update $CH(t)$ by finding the common tangent with the new curve $\wrb (xp_i)$. Using a suitable data structure for $CH(t)$, this can be done in $O(\log n)$ time. Then we check whether $\wrb (xp_i)$ intersects $CH(\leftn (t))$, which takes logarithmic time as well.
If it doesn't, we simply let $\rightn (t)$ be the current node. Otherwise $CH(\rightn (t))$ is completely masked by $\wrb (xp_i)$, since it lies below the intersection (see Figure~\ref{fig:boundaryCHupdate}). Thus we mark 
 $\rightn (t)$ as empty, thereby deleting the whole right subtree. We let $\leftn (t)$ be the current node. 
 
 When we are finished, the right subtrees corresponding to a whole ``prefix" of $\wrb$ are deleted, and $T$ is a smaller version of the exponential form described in Figure~\ref{fig:exptree}.

What remains is to insert the new leaf corresponding to $\wrb (xp_i)$ at the correct location in $T$, preserving the structure of the tree. Note that in the ray shooting query algorithm, if one of the two children of a node is marked empty, we directly jump to the valid node. All convex hull updates and intersection tests take $O(\log n)$ time, and the total number of nodes traversed is $O(\log n)$ as well, yielding $I(n) = O(\log^2 n)$ worst-case running time.

Since the worst-case complexity is $O(\log^2 n)$, rollbacks can be implemented by memorizing all updates, and performing them in reverse order. This causes the space complexity to increase to $O(n\log n)$.

Our algorithm only outputs the list of clusters. Section~\ref{sec:cloud} deals with the problem of identifying whether the highway is used for at least one pair of points. If this situation does not hold, we simply compute the convex hull of the set of points. Otherwise, we compute the convex hulls of the clusters, and join them to the highway with vertical segments. Details of this procedure can be found in~\cite{P03,YL07}. As previously mentioned, the highway hull may have open boundaries. To determine if an edge on the convex hull of a cluster is part of the highway hull, we check if $H$ is used between its endpoints (we simply compare the time using the highway or not between these two points).

\begin{lemma}
The Euclidean highway hull can be computed in $O(n\log^2 n)$ worst-case time using $O(n \log n)$ space when the speed $v$ is infinite.
\end{lemma}

\subsection{Finite speed}
\label{sec:Bounded}
When $v$ is bounded, the walking region of a point consists of a pair of half-parabolas tangent to $H$. To generalize the infinite-speed algorithm,
it suffices to verify that all the key properties used in the infinite speed version still hold when $v$ is bounded. 
Yu and Lee~\cite{YL07} provide many of the required details to generalize the algorithm.
In particular, the shape of the clusters must be adapted to take into consideration the fact that shortest paths enter $H$ 
at some angle $\alpha < \pi/2$ instead of orthogonally. 

Concerning
the algorithm itself and the identification of clusters, nothing changes. The properties of monotonicity~\cite[Lemma 2.1.11]{P03} of the right boundary of the clusters, as well as the ordering of the intersections between the boundaries of clusters and a vertical ray~\cite{YL07} are preserved. For the new steps introduced in our algorithm (namely the ray shooting data structure for segment queries), our reasoning only uses the assumption of monotonicity and thus remains valid in the general setup.

Once convex hulls are computed, they are to be joined to $H$ with segments whose direction obey Snell's law of refraction. Detecting if the highway hull is open or closed can be handled as before.

\begin{theorem}
The Euclidean highway hull can be computed in $O(n\log^2 n)$ worst-case time using $O(n \log n)$ space.
\end{theorem}

\section{Useful Highways}
\label{sec:cloud}

Let $H$ be a highway such that $\HH_2(S,H)$ differs from the standard convex hull $CH(S)$,
for a fixed speed $v>1$. Now consider translating $H$
continuously to infinity; it is clear that when $H$ is far enough
from the point set both hulls coincide:
$\HH_2(S,H) = CH(S)$. 
More generally, if we
consider the set of all lines that are parallel in any given
direction, we obtain a strip of \emph{useful} lines and two
halfspaces of \emph{useless} lines, in the sense that
no subset of a useless line serves as part of a shortest path between any points in $S$.

A highway on the bounding line $L$ of the strip can be used by at least one pair
of points, as part of a path that is equally short as the line segment between the points.

\begin{figure}[htbp]
 \begin{center}
 \includegraphics[scale=0.75]{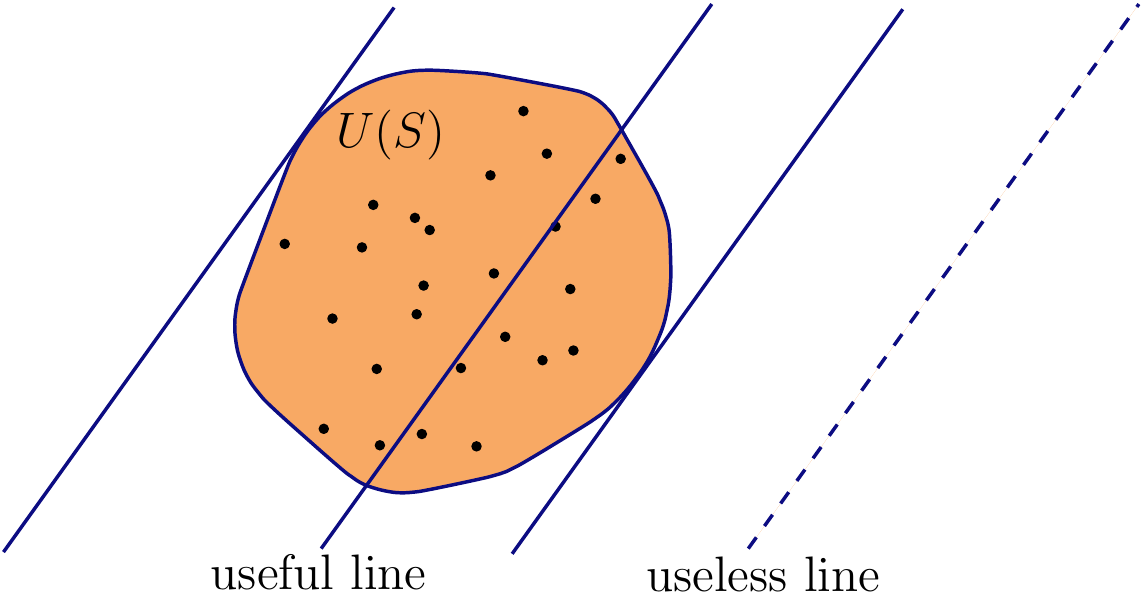}
 \end{center}
 \caption{The useful region of a point set $S$.}
 \label{fig:useful}
\end{figure}

If we repeat this construction for all possible directions, we obtain
the \emph{useful region} for $S$, denoted
by $U(S)$ (see Figure~\ref{fig:useful}): $U(S)$ is the locus of the points in the plane such that if a highway does not intersect $U(S)$, this highway is not used for any pair in $S$. Hence, a line is useless
as a highway if and only if it does not intersect $U(S)$. The lines
supporting $U(S)$ are precisely those that bound the strip of
useful lines in some direction.

\begin{figure}[htbp]
 \begin{center}
 \includegraphics[scale=0.75]{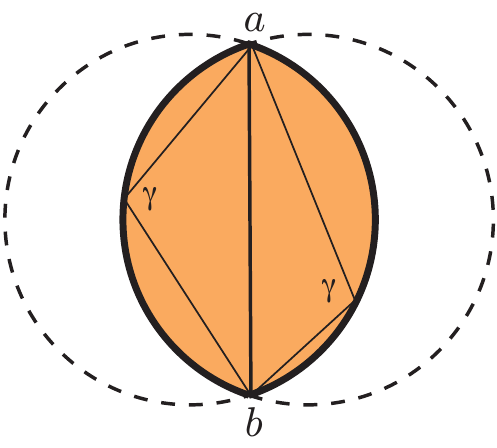}
 \end{center}
 \caption{The lens $\ell_{\gamma}(a,b)$.}
 \label{fig:lens}
\end{figure}

For two points $a$ and $b$, denote by $\ell_{\gamma}(a,b)$ the symmetric
\emph{lens} composed of two circular arcs of equal radii joined at $a$ and
$b$, such that the segment $ab$ is seen from every point on the
boundary with aperture angle $\gamma$ (see Figure~\ref{fig:lens}).

For a given speed $v>1$ we denote
by $\alpha$ the angle such that $\sin \alpha=1/v$.
\begin{lemma}\label{lem:usefulForTwo}
 The useful region for the point set $\{a,b\}$ is
$U(\{a,b\})=\ell_{\alpha+\pi/2}(a,b)$.
\end{lemma}
\begin{proof} Let $H$ be a line tangent to the lens
$\ell_{\alpha+\pi/2}(a,b)$ at point $c$. We wish to prove that the time required
for travelling from $a$ to $b$ using $H$ as a highway with speed
$v$ is equal to the Euclidean distance $|ab|$ (refer to
Figure~\ref{fig:cloudForTwo}).

\begin{figure}[htbp]
 \begin{center}
 \includegraphics[scale=0.75]{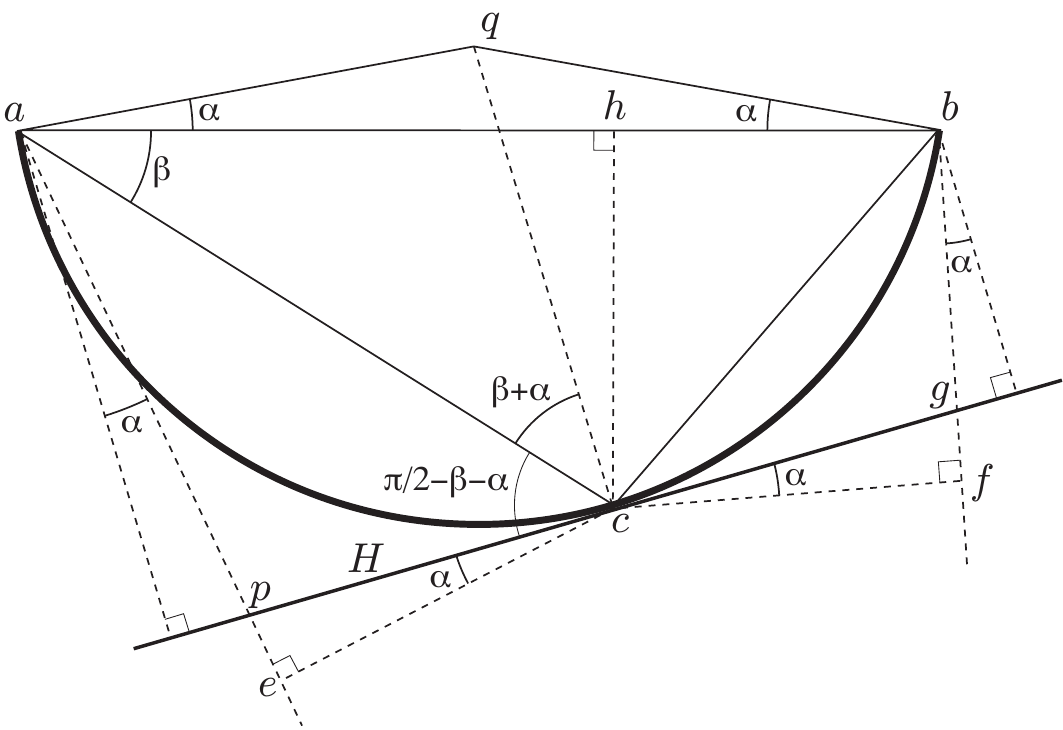}
 \end{center}
 \caption{Proof of Lemma~\protect\ref{lem:usefulForTwo}.}
 \label{fig:cloudForTwo}
\end{figure}

A counterclockwise rotation of
the line through $a$ perpendicular to $H$
by angle $\alpha$ yields a
line that crosses $H$ at a certain point $p$. 
Similarly, a clockwise rotation of
the line through $b$ perpendicular to $H$
by angle $\alpha$ yields a line that crosses
$H$ at a certain point $g$.

Construct the lines through $c$ perpendicular to $ap$ and $bg$, and let
$e$ and $f$ be their respective intersections with the lines through $ap$ and $bg$. By construction,
$\widehat{ecp}=\widehat{gaf}=\alpha$.

Let $\beta=\widehat{bac}$. The triangle
$\triangle cqa$ is isosceles, therefore
$\widehat{acq}=\widehat{qac}=\alpha+\beta$. As the lines $qc$ and
$H$ are perpendicular, we have $\widehat{pca}=\pi/2-\alpha-\beta$,
and hence
$\widehat{eac}=\pi/2-[(\pi/2-\alpha-\beta)+\alpha]=\beta$.
Therefore, if we denote by $h$ the intersection point of the line through $c$
perpendicular to $ab$, we see that triangles $\triangle hca$
and $\triangle eca$ are congruent, which implies
that $|ae|=|ah|$.

Taking into account that $|ae|=|ap|+|pe|$ and that
$\sin\alpha=1/v$, we obtain
\begin{equation}\label{eq:left}
|ah|=|ae|=|ap|+|pc|/v
\end{equation}

An identical argument leads to
\begin{equation}\label{eq:right}
|hb|=|bf|=|bg|+|gc|/v
\end{equation}
and from equations (\ref{eq:left}) and (\ref{eq:right}) we
obtain
$$
|ab|=|ap|+|pg|/v+|gb|
$$
which means that the Euclidean distance between $a$ and $b$ is
equal to the time required for traveling from $a$ to $b$ using $H$, as claimed.
\qed\end{proof}

\begin{lemma}\label{lem:usefulForMoreThanTwo}
The useful region for a point set $S$ is $U(S)= CH(\cup_{a,b\in S} U({a,b}))$.
\end{lemma}

\begin{proof}
Let $R=\cup_{a,b\in S}\ell_{\alpha+\pi/2}(a,b)$ be the union of the
$n\choose 2$ lenses. By Lemma~\ref{lem:usefulForTwo}, a line $H$ is useful if
and only if it intersects at least one of the lenses. Equivalently, $H$ is
useful if and only if $H \cap CH(R)\neq\varnothing$.
\qed\end{proof}

Lemma~\ref{lem:usefulForMoreThanTwo} gives a brute-force algorithm for
the computation of $U(S)$: For every pair of points $a,b\in S$,
compute $\ell_{\alpha+\pi/2}(a,b)$; then
$U(S)=CH(R)$ can easily be computed in $O(n^2 \log n)$ time.
It is clear that the $n\choose 2$ lenses are not independent, as
they arise from a set of $n$ points. Thus we proceed to reduce the running time:

\begin{theorem}\label{lem:usefulRegion}
Let $S$ be a set of $n$ points in the plane, and let $v>1$ be any
given speed. Then the useful region $U(S)$ can be computed in
$O(n\log n)$ time.
\end{theorem}
\begin{proof} Notice that $U(S)\supset CH(S)$, because
$U(S)=CH(\cup_{a,b\in S}\ell_{\alpha+\pi/2}(a,b))\supset CH(S)$.
If $z\in U(S)$ is outside $CH(S)$, then
there is a pair of points $a,b\in S$ such that
$\widehat{bza}\ge\alpha+\pi/2$. This implies that there is
also a pair of vertices $p,q$ of $CH(S)$ such that
$\widehat{qzp}\ge\alpha+\pi/2$. Therefore, in order to obtain
$U(S)$ it suffices to construct $cloud_{\alpha+\pi/2}(CH(S))$, the
set of points that see $CH(S)$ with aperture angle $\alpha+\pi/2$,
because then $U(S)=CH(cloud_{\alpha+\pi/2}(CH(S)))$ (see
Figure~\ref{fig:cloud}).

\begin{figure}[htbp]
 \begin{center}
 \includegraphics{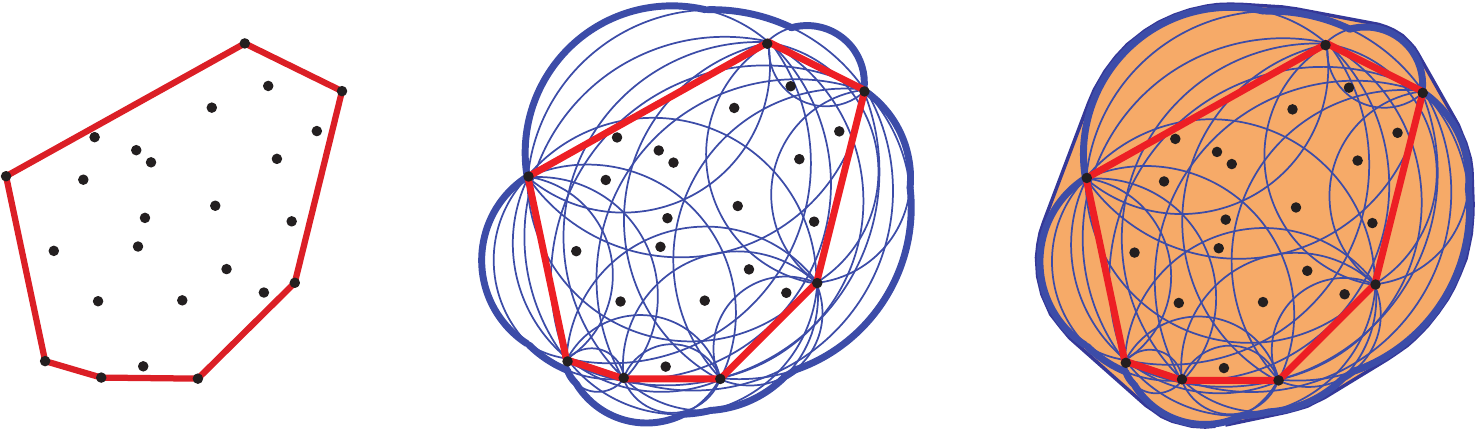}
 \end{center}
 \caption{Constructing the useful region of a point set $S$, for $v=1/\sin(20^\circ)$: Left: $S$ and $CH(S)$. Center: $cloud_{\alpha+\pi/2}(CH(S))$. Right: $U(S)$.}
 \label{fig:cloud}
\end{figure}

This cloud computation can be accomplished by taking a wedge of
aperture $\alpha+\pi/2$ supporting $CH(S)$ and using the rotating
callipers technique, as described in
\cite{snapshots,thesisTeichman}. The cloud consists of at most $2n$
circular arcs (one for every time that an arm of the rotating
wedge is flush with an edge of the polygon), and is obtained in
$O(n)$ time once $CH(S)$ is available. Notice that the region enclosed
by the cloud is star-shaped from any point inside $CH(S)$, and remains that way when
we bridge consecutive arcs by their common tangent bridge. Therefore,
once the cloud has been constructed, its convex hull $U(S)$ is easily obtained
via divide and conquer in $O(n\log n)$ time.
\qed\end{proof}

\section{Discussion}
We provided a $\Theta(n \log n)$ algorithm for the orthogonal highway hull.
If the input is a set of points sorted along the direction of the highway, 
our algorithm takes only $\Theta(n)$ time.
In the Euclidean case, we improved the previous $O(n^2)$ algorithm in~\cite{HPS99,P03} to $O(n\log^2 n)$, but it remains open whether this is optimal. 
We believe that by slightly modifying our approach and using fractional cascading, we could save a logarithmic factor. A lower bound of $\Omega(n\log n)$ is easily deduced from the fact that the output of the algorithm is $CH(S)$ if the highway is not useful.

A natural extension of this work would be to adapt these results to more \emph{realistic} highways and road networks. What if the highway is represented by a line segment (rather than an infinite line)? What if we have a highway network, with multiple highways crossing one another? 

\bibliography{highwayhull.bib}
\bibliographystyle{plain}

\end{document}